\documentclass[acmsmall,screen,preprint]{acmart}

\usepackage{soul}
\usepackage{graphicx}
\usepackage{xcolor}
\usepackage{tikz}
\usepackage{fontawesome5}
\usepackage{adjustbox}
\usepackage{multirow}
\usepackage{placeins}
\usepackage{enumitem}
\usepackage{subcaption}
\usepackage{colortbl}
\AtBeginDocument{}

\newcommand{\subr}[1]{{\small\texttt{r/#1}}}

\newcommand{\rev}[1]{\textcolor{black}{#1}}

\definecolor{none}{HTML}{e6e6e6}
\definecolor{adaptation}{HTML}{fdbf6f}
\definecolor{adaptation-header}{HTML}{ff7f00}
\definecolor{personalization}{HTML}{a6cee3}
\definecolor{personalization-header}{HTML}{1f78b4}
\newcommand{\Ba}{\colorbox{none}{\texttt{Ba}}}
\newcommand{\Mu}{\colorbox{none}{\texttt{Mu}}}
\newcommand{\Hs}{\colorbox{none}{\texttt{Hs}}}
\newcommand{\Redd}{\colorbox{adaptation}{\texttt{Re}}}
\newcommand{\Prev}{\colorbox{adaptation}{\texttt{Pr}}}
\newcommand{\Summ}{\colorbox{personalization}{\texttt{Su}}}
\newcommand{\Hist}{\colorbox{personalization}{\texttt{Hi}}}

\newcommand*\emptydot[1][0.66ex]{\tikz\draw (0,0) circle (#1);} 
\newcommand*\fulldot[1][0.66ex]{\tikz\fill (0,0) circle (#1);}

\begin{document}

\title{Contextualized Counterspeech Can Be More Persuasive Than Generic Counterspeech}

\author{Lorenzo Cima}
\email{lorenzo.cima@phd.unipi.it}
\affiliation{\institution{University of Pisa and IIT-CNR}
  \country{Pisa, Italy}
}

\author{Alessio Miaschi}
\email{alessio.miaschi@ilc.cnr.it}
\affiliation{\institution{ILC-CNR}
  \country{Pisa, Italy}
}

\author{Amaury Trujillo}
\email{amaury.trujillo@iit.cnr.it}
\affiliation{\institution{IIT-CNR}
  \country{Pisa, Italy}
}

\author{Marco Avvenuti}
\email{marco.avvenuti@unipi.it}
\affiliation{\institution{University of Pisa}
  \country{Pisa, Italy}
}

\author{Felice Dell'Orletta}
\email{felice.dellorletta@ilc.cnr.it}
\affiliation{\institution{ILC-CNR}
  \country{Pisa, Italy}
}

\author{Stefano Cresci}
\email{stefano.cresci@iit.cnr.it}
\affiliation{\institution{IIT-CNR}
  \country{Pisa, Italy}
}

\renewcommand{\shortauthors}{Cima et al.}

\begin{abstract}
AI-generated counterspeech offers a scalable and effective strategy to mitigate online toxicity by promoting more constructive dialogue. Yet, existing approaches adopt a generic, one-size-fits-all paradigm, overlooking the conversational context and characteristics of the targeted users. Here, we propose and evaluate multiple strategies for generating contextualized counterspeech that is \textit{adapted} to the moderation setting and \textit{personalized} to the moderated user. In detail, we explore a range of configurations that integrate different forms of contextual information and fine-tuning techniques. We conduct a comprehensive evaluation combining quantitative indicators with a pre-registered, mixed-design crowdsourcing experiment. To ensure robustness, we implement algorithmic measures of counterspeech quality based on ROUGE, BLEU, and BERTScore, observing overall consistent results across metrics. Furthermore, we analyze which characteristics of both the generated counterspeech and the moderated toxic message most strongly influence perceived persuasiveness, yielding insights into how contextualized interventions can be made more effective. 
\rev{Our findings show that personalization can be effective, but not uniformly so. Lightweight strategies combining conversational context and user history improve perceived adequacy and persuasiveness, whereas several other contextualization strategies degrade human-perceived counterspeech quality.}
Taken together, these results provide actionable directions for developing more personalized, effective, and responsible counterspeech systems, ultimately advancing human-AI collaboration in online content moderation.

\noindent {\small\faExclamationTriangle}~~\textbf{Warning:} \textit{This paper contains examples that may be perceived as offensive or upsetting. Reader discretion is advised.}
\end{abstract}



\received{20 February 2007}
\received[revised]{12 March 2009}
\received[accepted]{5 June 2009}

\maketitle

\section{Introduction}
\label{sec:introduction}
Online toxicity encompasses hateful, offensive, or otherwise harmful language that can cause emotional distress or push users to disengage from digital conversations. Its consequences are both social and economic: online, toxicity hampers participation, disrupts healthy discourse, and exacerbates divisions~\cite{aleksandric2024users}. \rev{Offline, it has been linked to physical violence, diminished respect for norms, and significant psychological harm~\cite{gallacher2021online}. These risks are also difficult to anticipate from user activity alone, as harmful trajectories may not be easily predictable from observable online behavior~\cite{cerulli2026dark}. As a result, tackling online toxicity has become a pressing issue for both regulators and platform operators~\cite{shahi2025year}.}

\begin{figure}[t]
\centering
    \includegraphics[width=0.7\columnwidth]{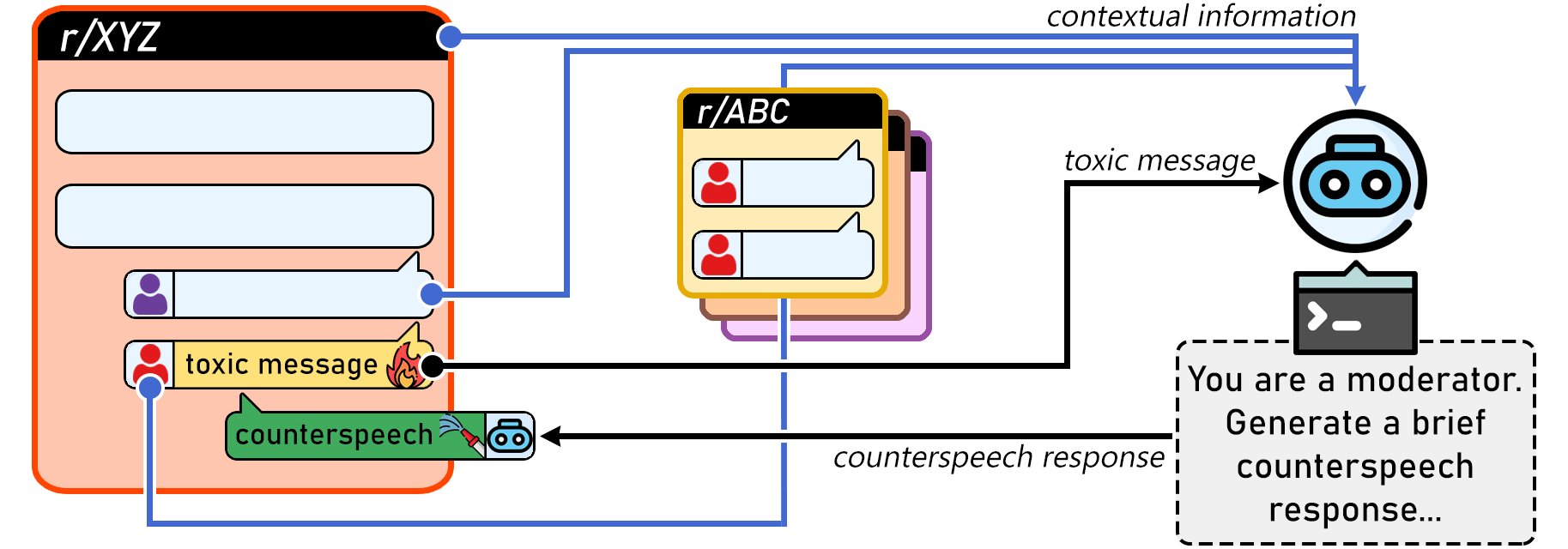}
    \caption{Current AI-generated counterspeech primarily relies on the content of the toxic message alone. In contrast, we generate contextualized counterspeech that integrates information about the community, the surrounding conversation, and the moderated user, aiming to produce more persuasive and targeted responses.}
    \label{fig:intro-example}
\end{figure}

To mitigate toxicity and other online harms~\cite{trujillo2025dsa,dwork2024content,cresci2014criticism}, platforms employ a range of moderation tools, including user bans, content removals, demotion, and friction-based interventions~\cite{trujillo2022make,chandrasekharan2022quarantined,tessa2025beyond}. One promising alternative to top-down moderation is counterspeech—user-driven replies aimed at de-escalating or challenging toxic content~\cite{garland2022impact,bonaldi2024nlp,tibau2025prevalence}. Counterspeech offers the advantage of avoiding censorship and deterring toxic users from simply migrating to other spaces~\cite{horta2021platform}. However, several factors limit its scalability and effectiveness. Manual counterspeech can place a significant emotional burden on users who regularly confront hostility~\cite{steiger2021psychological}, and it also raises safety concerns due to the risk of retaliation~\cite{tabassum2024investigating}.
Moreover, the sheer scale of toxic content makes manual responses unfeasible. In response, researchers have turned to automated counterspeech generated by large language models (LLMs)\cite{tekirouglu2020generating,bonaldi2024nlp}. \rev{However, evaluating the true effectiveness of such systems remains difficult~\cite{gillespie2020content,zheng2026validating}. Prior evaluations have often focused on surface-level properties, such as fluency or grammaticality, while deeper dimensions such as contextual fit, persuasive impact, and user-specific effectiveness remain harder to assess~\cite{zubiaga2024llm,kumar2025counterspeech}. Recent work has advanced automatic and LLM-as-a-judge evaluation frameworks for counterspeech~\cite{hengle2025cseval,ngueajio2025think}, but human-centered validation of contextualized and personalized counterspeech remains limited.}
A key shortcoming is that most AI-generated counterspeech is one-size-fits-all: responses are based only on the toxic message, ignoring critical contextual elements like the conversation history, the community norms, or the characteristics of the toxic user~\cite{cresci2022personalized}. Yet, effective moderation requires context-sensitive responses~\cite{gillespie2020content,huang2024opportunities}, calling for a shift from generic to tailored interventions~\cite{cresci2022personalized}.

\paragraph{Scientific focus} This study extends our previous work~\cite{cima2025contextualized} to develop and evaluate a method to generate contextualized counterspeech that is \textit{adapted} to the community and conversation context, and \textit{personalized} to the individual being moderated. Unlike traditional approaches, our system goes beyond the toxic message and integrates broader information about the surrounding environment (community, moderated user, and specific conversation), as illustrated in Figure~\ref{fig:intro-example}. We test whether contextualized responses are more persuasive than generic ones, conducting experiments in politically active Reddit communities.
We identify what makes counterspeech effective and design both algorithmic and human evaluations to measure these qualities. For aspects not adequately captured by existing metrics~\cite{zubiaga2024llm}, we conduct a pre-registered, mixed-design crowdsourced user study.
Our findings highlight both the promise and the complexity of generating contextualized counterspeech. In summary, our main contributions are:

\begin{itemize}[leftmargin=*]
    \item We propose and test novel strategies for generating \textit{adapted} and \textit{personalized} counterspeech that accounts for the community, user, and conversation context.
    \item \rev{We show that incorporating contextual information can enhance perceived adequacy and persuasiveness for specific lightweight prompting configurations, while also demonstrating that many adaptation and personalization strategies do not improve over generic baselines and can even degrade human-perceived quality.}
\item We identify which characteristics of the toxic message and the counterspeech (i.e., subtypes of toxicity and emotional profile) influence human perceptions of persuasiveness, providing guidelines for future developments.
    \item\rev{We show that known limitations of automated evaluation metrics also apply to contextualized counterspeech, underscoring the need for evaluation frameworks that better capture human judgments of counterspeech quality.}
\end{itemize}
 \section{Related Work}
\label{sec:related_work}

\subsection{LLM-generated counterspeech}
Counterspeech is a moderation strategy that mitigates harmful online behavior without relying on restrictive actions, thereby preserving freedom of expression~\cite{chung2023understanding}. Prior work has shown its potential across observational, quasi-experimental, and controlled experimental settings, including interventions by scholars, NGOs, and individual users~\cite{chung2021towards,goffredo2022counter,hassan2023discgen,garland2022impact}. However, its effectiveness depends on message design, target audience, platform context, and intervention strategy, and there is still limited understanding of which types of counterspeech work best under which conditions~\cite{chung2023understanding,hickey2026assessing}.
LLMs offer a scalable way to generate counterspeech, and recent work has explored several strategies to improve response quality. \citet{tekiroglu2022using} evaluate pre-trained LLMs and decoding strategies for generating persuasive counterspeech. \rev{\citet{wang2024f2rl} instead use reinforcement learning to optimize counterspeech for factuality and faithfulness.} Other approaches incorporate socio-psychological strategies, such as empathy, disapproval, appeals to moral norms, warnings about consequences, and humor~\cite{bar2024generative,gennaro2025counterspeech}. Further work uses Retrieval-Augmented Generation to improve factual grounding and specificity~\cite{jiang2025rezg,leekha2024war,anik2025multi}, or emotional targeting methods, such as attribute prefix learning, to adapt the affective tone of the response~\cite{kumar2025counterspeech}.

Despite these advances, most LLM-based counterspeech systems generate responses mainly from the toxic message itself, with limited attention to the broader conversational setting or to the characteristics of the target user. Among the few exceptions are \citet{dougancc2023generic} who incorporate demographic attributes, \citet{bar2024generative} who generate counterspeech conditioned on the hateful post, and \citet{ngueajio2025think} who use persona-based LLMs. Our work extends this nascent line of research~\cite{cresci2022personalized} by systematically comparing multiple contextual signals, including user profiles, conversational dynamics, and community norms, and by evaluating how these signals affect perceived counterspeech quality.

\subsection{Persuasiveness of LLM-generated messages}
The persuasive potential of LLMs has been studied across several domains, from LLM-to-LLM persuasion to human attitude and behavior change~\cite{jones2026lies,breum2024persuasive,salvi2025conversational}. Much of this work focuses on whether generated messages can influence users' decisions or beliefs. For example, \citet{furumai2024zero} compare emotional appeals, rational arguments, and stylistic variations in donation campaigns, while \citet{karande2024persuasion} show that LLMs can act as virtual sales agents by adapting language to customer profiles. In a more adversarial setting, \citet{goldstein2024persuasive} show that AI-generated propaganda can be as persuasive as human-written one. Political and social persuasion have also received increasing attention. \citet{hackenburg2025scaling} study how LLM-generated messages affect political attitudes, voter preferences, and civic engagement, with evidence that larger models do not necessarily produce stronger persuasive effects. Other studies examine belief change in more deliberative settings. \citet{costello2024durably} use LLM-based dialogues to reduce conspiracy beliefs, finding effects that persist over time, whereas \citet{potter2024hidden} show that biased LLMs can unintentionally shape users' opinions even when they are not explicitly designed to persuade.
Personalization is a recurring factor in this literature. \citet{pauli2024measuring} examine how persona-based variations affect message persuasiveness across audiences, while \citet{gennaro2025counterspeech} apply social-psychological framing strategies to generate persuasive counterspeech aimed at reducing intergroup hostility. Relatedly, LLMs have been explored as moderators or facilitators of online discourse, with studies evaluating their ability to mitigate toxicity, depolarize discussions, or improve conversational outcomes through constructive responses~\cite{cho2023can,hong2024outcome,govers2024ai,hickey2026assessing}. However, the persuasiveness of contextualized and personalized counterspeech remains comparatively underexplored, especially when considering how different contextual signals affect perceived adequacy, persuasive potential, and user-specific fit.

\subsection{LLM adaptation and personalization}
Most moderation strategies follow a one-size-fits-all approach, applying the same intervention across users and contexts. While scalable, this ignores the psychological and social dynamics underlying harmful behavior. Recent work therefore highlights the potential of contextualized and personalized interventions~\cite{trujillo2023one,costello2024durably,cima2025investigating,cresci2022personalized}. LLMs offer a scalable way to adapt messages to users and situations, but their use for personalized moderation and counterspeech remains limited.
Some work aligns counterspeech with the toxic message or its affective tone. \rev{Most closely related to our work, Bär et al.~\cite{bar2024generative} tested LLM-generated contextualized counterspeech in a field experiment, finding that it could backfire by increasing toxicity, while generic warning-based counterspeech reduced subsequent hateful posting. This result shows that contextualization is not necessarily beneficial by itself and that more systematic work is needed to understand which design choices make contextualized counterspeech effective.} Similarly, \citet{kumar2025counterspeech} adapt counterspeech to emotional tone using emotion-guided prefix learning.
Other personalization strategies adapt LLM outputs to user-level traits. Some works condition messages on socio-demographic characteristics~\cite{salvi2025conversational,dougancc2023generic,beck2024sensitivity,giorgi2024human}, while others use personality traits or persona-based profiles to shape the tone, framing, and argumentative style~\cite{jiang2023personallm,ngueajio2025think,pauli2024measuring}. In the political and disinformation domains, LLMs have also been used to tailor persuasive messages to political attitudes, prior beliefs, or target-group characteristics~\cite{hackenburg2024evaluating,zugecova2024evaluation,borah2026persuasion}. These studies suggest that personalization can affect persuasive or corrective interventions by improving audience fit, but they also raise concerns about behavioral targeting, manipulation, and unequal treatment across user groups. 

 \section{Problem definition}
\label{sec:problem}
$\mathbf{T} = \langle m_0, m_1, \ldots, m_N\rangle$ represents an online conversation thread, where messages $m_0, m_1, \ldots, m_N$ appear in chronological order. Given a toxic message $m_i$, our goal is to generate a counterspeech response $\hat{m}_{i+1} = \mathcal{G}(m_i, \mathbf{C}_i)$, where $\mathcal{G}$ is the counterspeech generator and $\mathbf{C}_i$ denotes the contextual information. Unlike prior work, where $\mathcal{G}$ only takes $m_i$ as input~\cite{cho2023can,he2023reinforcement,hong2024outcome,leekha2024war,bar2024generative}, here we generate \textit{adapted} and \textit{personalized} counterspeech by also incorporating $\mathbf{C}_i$ as input to $\mathcal{G}$.

Next, we identify two sets of properties that define high-quality counterspeech. The first includes fundamental goals of a counterspeech, while the second includes properties that are specific to AI-generated contextualized content. We intentionally exclude basic linguistic features like fluency and grammaticality, which are already reliably handled by modern LLMs~\cite{li2024pre}. \\

\subsection{Desired properties of effective counterspeech}
\begin{itemize}[leftmargin=*]
    \item \textit{Politeness.} Polite messages are more likely to engage toxic users and bystanders constructively~\cite{yu2024hate}. Politeness also aligns with ethical standards by discouraging hostile interactions.
    \item \rev{\textit{Adequacy.} Effective counterspeech should be suitable as an intervention to the toxic message, appropriately addressing or challenging the harmful content~\cite{bonaldi2024nlp}.}
\item \textit{Relevance.} Counterspeech must align with the specific content and context of the thread. Generic replies often fall short in terms of resonance and impact~\cite{gillespie2020content,cresci2022personalized}.
    \item \textit{Diversity.} A diverse set of responses can appeal to a broader range of users and situations~\cite{lees2022new}, while also improving perceived authenticity and genuineness, avoiding predictable responses.
    \item \textit{Truthfulness.} Accurate and factual responses are essential to preserving credibility and trust within the community~\cite{bonaldi2024nlp,wang2024f2rl}.
    \item \textit{Persuasiveness.} Persuasive counterspeech encourages attitude or behavior change, either by directly influencing the toxic user or by promoting prosocial norms among bystanders to have positive interactions~\cite{hong2024outcome}.
\end{itemize}

\subsection{Relevant properties of contextualized AI-generated counterspeech}
\begin{itemize}[leftmargin=*]
    \item \textit{Adaptation.} Counterspeech should reflect the norms, tone, and content of the specific community or conversation. Adapted messages are more likely to be accepted and taken seriously, increasing their credibility and persuasive power~\cite{gillespie2020content}.
    \item \textit{Personalization} focuses on user characteristic and behavior, differently from adaptation that is based only on the conversation context. Tailoring counterspeech to the traits or history of the toxic user can enhance rapport, reduce resistance, and increase impact~\cite{cresci2022personalized}. Personalization strengthens persuasion by showing empathy or strategic alignment.
    \item \textit{Artificiality} refers to the degree to which counterspeech is perceived as being machine-generated rather than human-authored. A high level of perceived artificiality can undermine the message’s credibility and emotional resonance. Moreover, perceived artificiality can trigger negative reactions, especially if users feel deceived or dismissed by automated moderation tools~\cite{goel2024artificial}. \rev{Recent work further shows that AI-generated and human-written counterspeech can be distinguishable by both humans and classifiers, partly due to differences in linguistic characteristics, politeness, and specificity~\cite{song2025assessing}.} Reducing artificiality by improving the naturalness, tone, and specificity of the message, helps counterspeech feel more genuine, increasing the likelihood that it will be taken seriously and achieve its persuasive intent~\cite{gillespie2020content}.
\end{itemize}
 \section{Methods}
\label{sec:method}
\subsection{Generation}
\label{sec:method-generation}

We employ an instruction-tuned version of LLaMA2-13B\footnote{\url{https://huggingface.co/dfurman/Llama-2-13B-Instruct-v0.2}} for generating counterspeech responses. The instruction prompts used during generation are listed in Appendix~\ref{appendix:prompts}. Then, we explore a range of generation configurations by varying the input information provided to the model and the data used for fine-tuning.\footnote{All models are publicly available at \url{https://huggingface.co/collections/alemiaschi/contextualized-counterspeech-llama-2-models-679e322e663f033e1aa654f2}} Below, we describe the experimental factors under evaluation, each identified by a unique [\texttt{label}]. Some factors do not involve adaptation nor personalization:
\begin{itemize}[leftmargin=*]
\item \textbf{Base} [\Ba]: The unmodified instruction-tuned LLaMA2-13B model. When used alone, this constitutes our \emph{baseline} configuration.
\item \textbf{Counterspeech fine-tuning}: We fine-tune the base model for the task of counterspeech generation using two publicly available datasets: 
[\Mu] \textsc{MultiCONAN}~\cite{fanton-etal-2021-human} which comprises 500 curated pairs of hate speech and counterspeech spanning multiple target categories (e.g., race, religion, nationality, sexual orientation, disability, and gender); 
[\Hs] the Reddit hate-speech intervention (RHSI) dataset~\cite{qian-etal-2019-benchmark} consisting of 5,020 Reddit conversations containing human-authored interventions. For consistency with our generation task, we retain only instances featuring a toxic comment paired with a human-written reply, and with a maximum message length of 250 words, yielding a filtered subset of 2,974 examples.
\end{itemize}
\rev{The above configurations provide reference settings inspired by prior work on automatic counterspeech generation, where responses are generated from the toxic message alone and models are optionally fine-tuned on counterspeech or intervention datasets~\cite{tekiroglu2022using,fanton-etal-2021-human}.}
In contrast, the following experimental factors introduce novel contextual information to the generator.

\subsubsection{Adaptation}
\label{sec:method-generation-adaptation}
To improve contextual alignment with the moderation setting, we introduce adaptation strategies that tailor the model’s output to the platform and conversation in which the toxic message occurs:
\begin{itemize}[leftmargin=*]
\item \textbf{Community} [\Redd]: As our study is situated within Reddit’s political communities, we adapt the generator to the platform’s stylistic and linguistic norms by fine-tuning it on comment-reply pairs sampled from five high-activity political subreddits (see Section~\ref{sec:dataset}). This adaptation familiarizes the model with Reddit-specific language, tone, and discourse conventions, making its responses more natural and platform-appropriate.
\item \textbf{Conversation} [\Prev]: Because toxic messages $m_i$ appear within broader threads, we enrich the model’s input with up to two preceding messages from the same conversation ($m_{i-1}, m_{i-2}$). This conversational context helps the model better understand the flow of discourse and produce counterspeech that directly responds to the ongoing exchange, rather than treating messages in isolation.
\end{itemize}

\subsubsection{Personalization}
\label{sec:method-generation-personalization}
To further increase the effectiveness of the counterspeech, we personalize the model’s responses based on user-specific characteristics. This allows the generator to produce interventions that are not only context-aware but also user-tailored: \begin{itemize}[leftmargin=*]
\item \textbf{Comment history} [\Hist]: We provide the generator with a window into the user’s prior behavior by prepending each toxic message $m_i$ with ten of the user's previous Reddit posts. This history offers implicit cues about the user’s tone, typical topics, and style, which the model can leverage to craft more targeted and relatable responses. 
\item \textbf{Summary} [\Summ]: As an alternative to providing raw message history, we distill a broader sample of the user's activity (twenty past messages) into a compact summary. This summary is generated by an instruction-tuned LLaMA2-13B model prompted to describe the user’s writing style, preferred lexicon, and main thematic interests (prompt detailed in Appendix~\ref{appendix:prompts}). These user profiles are then used as input to the counterspeech generator, serving as an explicit, structured source of personalization.
\end{itemize}

We implement and evaluate 36 different generation configurations by varying the combinations of the aforementioned factors.\footnote{For example, [\Ba\Prev\Hist] refers to the base LLaMA2-13B model receiving both prior conversation messages and the user's recent comment history as input.} \rev{The resulting design is factorial but not fully crossed. We excluded combinations that would be conceptually meaningless or structurally redundant. In particular, factors corresponding to different model variants are mutually exclusive: [\Ba] cannot be combined with [\Mu] or [\Hs], since these represent alternative model states rather than independent prompt-level factors. We also excluded combinations of factors that are mechanically dependent. For example, [\Summ] is derived from the same pool of prior user comments used for [\Hist]; combining the two would duplicate the same user information and make their individual effects difficult to interpret.}
When combining multiple datasets for fine-tuning (e.g., [\Mu] and [\Hs]), we adopt a multi-task learning setup by merging the datasets during training. This comprehensive design enables us to systematically analyze the contribution of each contextual and personalization factor to the quality of the generated counterspeech.

\subsection{Evaluation}
\label{sec:method-evaluation}
The goal of our evaluation is twofold: \textit{(i)} to assess the degree to which the generated counterspeech messages fulfill the desired properties outlined in Section~\ref{sec:problem}; and \textit{(ii)} to identify which generation configurations are most effective. To this end, we adopt a hybrid semi-automatic evaluation framework. First, we conduct a broad algorithmic assessment across all model configurations using a set of quantitative indicators. Then, based on this automatic screening, we select a subset of configurations for in-depth human evaluation.

\subsubsection{Algorithmic evaluation}
\label{sec:method-evaluation-algo}
Following recent work~\cite{bonaldi2024nlp,saha2022countergedi,he2023reinforcement}, we define a set of evaluation indicators, each corresponding to one or more of the targeted properties. These indicators allow us to systematically compare the outputs of different configurations:
\begin{itemize}[leftmargin=*]
    \item \textit{Relevance:} We assess the relevance as the topical alignment between each toxic message $m_i$ and its corresponding counterspeech $\hat{m}_{i+1}$ by computing the ROUGE score between the two texts, which captures lexical overlap and content similarity.
    \item \textit{Diversity:} To evaluate the variability of counterspeech responses within each configuration, we compute an intra-configuration diversity score:
    \[
        \text{Diversity} = 1 - \frac{1}{n(n-1)} \sum_{i=1}^{n} \sum_{\substack{j=1 \\ j \neq i}}^{n} \text{ROUGE}(\hat{m}_i, \hat{m}_j),
    \]
    where \( \text{ROUGE}(\hat{m}_i, \hat{m}_j) \) is the similarity between two counterspeech messages \( \hat{m}_i \) and \( \hat{m}_j \) and \( n \) is the number of generated messages in the configuration. Higher diversity indicates less repetition and more lexical variety.
    \item \textit{Readability:} We measure how readable the generated messages are using the Flesch Reading Ease Score (FRES), which is based on sentence length and word complexity.
    \item \textit{Toxicity:} To ensure that counterspeech does not inadvertently perpetuate harmful language, we evaluate the toxicity of each generated message using Google’s Perspective API.\footnote{\url{https://perspectiveapi.com/}} \rev{We use toxicity as a safety-oriented inverse proxy for impoliteness: while low toxicity does not necessarily imply that a response is polite, highly toxic counterspeech is incompatible with the goal of producing respectful and constructive interventions.}
    \item \textit{Adaptation:} We assess how much each configuration deviates from the baseline style and content by computing the diversity (i.e., $1 - \text{ROUGE}$) between the counterspeech messages generated by the baseline model [\Ba] and those produced by the configuration under evaluation. This captures the degree of contextual adaptation introduced.
    \item \textit{Personalization$_{\textnormal{lex}}$:} We measure how lexically aligned the counterspeech is with the moderated user’s typical language. Specifically, we compute the ROUGE score between each generated message $\hat{m}_{i+1}$ and a sample of prior messages authored by the same user.
    \item \textit{Personalization$_{\textnormal{wri}}$:} We also assess similarity in writing style by extracting linguistic profiles for both the counterspeech and the user messages using \textsc{ProfilingUD}, which captures over 130 syntactic and morphological features~\cite{brunato2020profiling}. We then compute the Spearman correlation between the stylistic profiles of each counterspeech message and the toxic user’s past writing.
\end{itemize}

\subsubsection{Configuration selection}
\label{sec:method-evaluation-selection}
While the algorithmic evaluation provides a scalable and systematic way to compare all configurations, it falls short in capturing subjective and nuanced properties such as \textit{persuasiveness} and \textit{artificiality}. Moreover, the accuracy and interpretability of some existing automatic indicators remain open to discussion~\cite{halim2023wokegpt,zubiaga2024llm}. To mitigate these limitations, we complement our quantitative analysis with a targeted human evaluation conducted via crowdsourcing.
Given the large number of configurations, a complete manual assessment is infeasible. Therefore, we adopt a principled approach to select a representative subset. First, we generate a performance ranking for each individual indicator by ordering all configurations according to their respective scores. These rankings are then aggregated into a single global ordering (i.e., a “super-ranking”) by solving an optimization task that minimizes Spearman’s footrule distance across the rankings. This super-ranking serves as the basis for configuration selection.
From the aggregated ranking, we select six representative configurations: the highest- and lowest-performing configurations among those that implement only adaptation, only personalization, and a combination of both. This selection strategy allows us to compare the effectiveness of each contextualization strategy across performance extremes, while also enabling us to evaluate the alignment between automatic indicators and human judgment. In addition, we include the baseline configuration [\Ba] in the evaluation set to serve as a reference point for all comparisons.
To ensure that the messages assessed by human annotators are representative of each configuration’s overall behavior, we further select a subset of counterspeech messages per configuration. For each of the seven selected configurations, we compute the centroid of the configuration across all evaluation indicators. We then identify and select the 20 counterspeech messages that are closest to this centroid in terms of their indicator vectors. This ensures that the messages chosen for human evaluation are not outliers, but rather typical examples of the outputs generated by each configuration.
This hybrid selection process ensures that the subsequent human evaluation is both meaningful and efficient, enabling a thorough assessment of the configurations’ real-world effectiveness in generating counterspeech.

\subsubsection{Human evaluation}
\label{sec:method-evaluation-human}

We conduct an extensive human evaluation through a pre-registered,\footnote{\url{https://aspredicted.org/b55z-5qy4.pdf}} mixed-design crowdsourcing experiment on Amazon Mechanical Turk.\footnote{This research received ethical approval from CNR's IRB (protocol \#0306210).} The study is structured around a two-tiered design comprising both between-subjects and within-subjects components. At the start of the task, participants are randomly assigned to one of two between-subjects conditions: \textit{(i)} a \textit{non-contextual} condition, in which participants are shown only the toxic message $m_i$ and the corresponding counterspeech $\hat{m}_{i+1}$; or \textit{(ii)} a \textit{contextual} condition, where participants are additionally shown the contextual information used to generate the counterspeech, including conversation history and user-specific features when applicable. After this initial assignment, participants proceed to evaluate multiple $(m_i, \hat{m}_{i+1})$ pairs in a within-subjects design. Each participant rates responses from all seven model configurations selected during the configuration selection phase. The presentation order of these configurations is randomized to control for order effects and potential fatigue biases. 
Participants are asked to evaluate each counterspeech response on a five-point Likert scale across several dimensions: its \textit{relevance} to the toxic message, its \textit{adequacy} as a form of counterspeech, its \textit{truthfulness}, its perceived \textit{artificiality}, and its \textit{persuasiveness}. \rev{We operationalize persuasiveness as \textit{perceived persuasiveness}, namely participants' third-party assessment of the response's persuasive potential. Specifically, following recent work~\cite{hong2024outcome}, persuasiveness is assessed through two distinct items: the perceived likelihood that the counterspeech (i) persuades the author of the toxic message to re-engage in a more civil manner, and that it (ii) steers the broader conversation back toward civil discourse.}
Participants in the \textit{contextual} within-subjects condition are also asked to assess how \textit{contextualized} the counterspeech feels in relation to the surrounding conversation and user behavior. Finally, all participants complete a brief socio-demographic questionnaire. The full list of evaluation questions and instructions is provided in Appendix~\ref{appendix:questionnaire}.
This mixed design enables a comprehensive analysis of both model performance and the role of contextual information. In particular, the between-subjects comparison isolates the added value of context in shaping users’ perceptions of counterspeech quality, while the within-subjects design allows for direct comparison across generation strategies.

\subsubsection{Statistical analysis}
To analyze the results from the within-subjects evaluation, we first assess whether differences across the evaluated configurations are statistically significant using Friedman tests. To identify specific configurations that differ significantly from the baseline, we perform pairwise comparisons using Wilcoxon's signed-rank tests, applying a Bonferroni correction to control for multiple hypothesis testing. For each comparison, we report effect sizes and confidence intervals using the matched-pairs rank biserial correlation coefficient.
In addition, we evaluate whether contextual information significantly influences participant judgments by comparing results between the two between-subjects groups. For each configuration, we perform two-sample Mann–Whitney U tests with Bonferroni correction, followed by effect size estimation via Glass rank biserial correlations. 

\subsubsection{Power analysis.} We target a sample size of  $\approx2,500$ participants for each within-subjects experiment. This number allows us to detect small effect sizes (Cohen’s $d = 0.2$) with strong statistical power (85\%) at a 95\% confidence level.  \section{Data}
\label{sec:dataset}

We collected Reddit comments posted over multiple years from five widely followed subreddits that regularly engage in discussions about U.S. politics. These include two ideologically polarized communities, one right-leaning (\subr{conservatives}) and one left-leaning (\subr{progressive}), as well as two subreddits centered around prominent political figures: \subr{the\_donald}, focused on Donald Trump (right-leaning), and \subr{aoc}, focused on Alexandria Ocasio-Cortez (left-leaning). To ensure a more ideologically mixed setting, we also included \subr{politics}---a large, general-interest political subreddit with varied viewpoints. To balance the volume of data across subreddits of different sizes, we collected comments spanning either 36 or 12 months per subreddit. For \subr{the\_donald}, collection ended in June 2020 when the subreddit was permanently banned~\cite{cima2025investigating,cerulli2026big}. All comments were retrieved using the Pushshift dataset~\cite{baumgartner2020pushshift}, which provides historical Reddit data.

\paragraph{Counterspeech dataset} To create a benchmark set of toxic comments for counterspeech generation, we first computed toxicity scores for all collected comments using Google's Perspective API~\cite{lees2022new}. \rev{We relied on Perspective API because it is widely used for measuring online toxicity and operationalizes toxicity as language that is likely to make someone leave a discussion, which is closely aligned with our goal of identifying comments that may disrupt civil conversation and motivate counterspeech interventions. We retained only those comments with a toxicity score $\geq 0.5$, a threshold commonly used in literature~\cite{trujillo2022make,kumarswamy2025causal}, and that were part of conversation threads with at least two parent messages. This filtering process yielded 292 toxic comments across 49 distinct threads. From this set, we discarded comments written by users with fewer than 20 comments in the considered period in all Reddit. In such cases we could not infer sufficient user-level information to generate personalized counterspeech, as required by the personalization strategies described in the next paragraph. This additional filtering step yielded the final benchmark of 128 toxic comments across 35 distinct threads, as shown in the Appendix Table~\ref{tab:dataset}. Although the number of threads is relatively balanced across the considered subreddits, most toxic comments originate from \subr{politics}. This pattern is consistent with the larger size and broader activity of \subr{politics} compared to the other communities in our dataset. At the same time, it highlights a realistic challenge for moderation systems: high-activity communities may generate a larger absolute volume of toxic content even when toxicity is distributed across multiple discussion threads. This further motivates the need for scalable counterspeech interventions that can help de-escalate toxic exchanges and support the restoration of constructive discussion.} Each of the 36 counterspeech generation configurations described in Section~\ref{sec:method-generation} was tasked with generating one counterspeech response for each of the 128 toxic messages, resulting in a total of 4,608 generated counterspeech responses to evaluate.

\paragraph{Adaptation and personalization datasets} To implement the contextual strategies outlined in Section~\ref{sec:method-generation}, we constructed additional datasets tailored to each type of adaptation or personalization. For the community adaptation factor [\Redd], we selected a random stratified sample of approximately 7,500 comment–reply pairs from the five political subreddits. This dataset was used to fine-tune the base model to the conversational norms, tone, and linguistic style typical of political discourse on Reddit. \rev{For the conversation-level adaptation factor [\Prev], we extracted the most recent parent message preceding each of the 128 selected toxic comments. Since our filtering procedure retained only toxic comments with available parent context, the dataset does not include top-level toxic comments. These parent messages were fed as contextual input via prompting at generation time. To implement personalization, we gathered twenty prior comments authored by each user who wrote one of the toxic messages, sampled from the user's activity across Reddit rather than restricted to the target subreddit. These samples were used to generate user summaries for the [\Summ] condition. Additionally, ten of these user-authored comments were directly prepended to the toxic comment during counterspeech generation to implement the comment history strategy [\Hist]. The selected comments were used as a practical proxy for the user's writing style, lexical choices, and recurring topics.} \section{Results}
\label{sec:results}

\subsection{Algorithmic evaluation}
\label{sec:algorithmic-results}
We evaluate all factors (\textit{N}=7) and configurations (\textit{N}=36) with the indicators defined in Section~\ref{sec:method-evaluation-algo}.

\subsubsection{Factors}
\rev{We begin our analysis with the modeling of indicator values based on configuration factors. Given that there is an overlap of factors among many of the 36 configurations and that all indicator values are between 0 and 1, we employed \emph{beta regression}, a robust approach to model beta distributions of values in the standard unit interval~\cite{cribari2010beta}. For every indicator, we used a traditional beta regression model---having the \emph{logit} link function and treating precision ($\phi$) as constant---based on all the 7 single factors and their 15 pairwise interactions used in the configurations. We did not consider higher-order interactions to avoid issues with model overspecification and the additional complexity. An overview of the results of the beta regressions is depicted in Table~\ref{tab:indicator-factor-betareg}, in which statistically significant ($p<.05$) estimates of means for factors are highlighted according to their relative effect on indicator values.} 

The analysis reveals that the effect of a factor is highly dependent on the specific dimension being evaluated. No factor produces across-the-board improvements. Instead, gains in some indicators often come at the cost of degradations in others. For example, fine-tuning on the \textsc{MultiCONAN} dataset [\Mu] and our Reddit-specific political conversations [\Redd] leads to noticeable improvements in \textit{relevance}, \textit{diversity}, and \textit{adaptation}~\cite{cima2025contextualized}. However, these same factors negatively impact \textit{toxicity} and  stylistic \textit{personalization}, either by themselves or when interacting with other factors, suggesting a trade-off between general linguistic alignment and user-specific adaptation. This trade-off underscores the complexity of the task: no single factor enhances all desired properties simultaneously. As such, practitioners deploying counterspeech systems may need to tailor configurations based on the specific requirements of a given use case—prioritizing readability or relevance in some scenarios, and personalization or toxicity in others. \rev{Some factors, however, perform poorly across most metrics. Fine-tuning on the RHSI dataset [\Hs] consistently degrades performance in terms of \textit{relevance}, \textit{diversity}, \textit{toxicity}, and writing style \textit{personalization}, which is sometimes exacerbated or alleviated by interacting with other factors}. 
\rev{On the \textit{personalization} front, [\Summ]  yields  improvements in both lexical and stylistic indicators when interacting with [\Hist], whereas [\Hist] degrades writing style and lexical \textit{personalization} by itself. Interestingly, the base factor [\Ba], which appears only in configurations without any fine-tuning, shows relatively favorable results in the stylistic \textit{personalization} indicator.} This suggests that fine-tuning on large-scale datasets such as \textsc{MultiCONAN}, RHSI, or even Reddit-specific conversations may dilute the model's ability to align with individual users’ writing patterns.

\begin{table}[t]
\fontsize{6.75pt}{7.0pt}\selectfont
\begin{tabular*}{\linewidth}{@{\extracolsep{\fill}}crrrrrrrrrrrrrr}
\toprule
 & \multicolumn{2}{c}{{\bfseries adaptation\,$\uparrow$}} & \multicolumn{2}{c}{{\bfseries diversity\,$\uparrow$}} & \multicolumn{2}{c}{{\bfseries perso.$_\text{lex}$\,$\uparrow$}} & \multicolumn{2}{c}{{\bfseries perso.$_\text{wri}$\,$\uparrow$}} & \multicolumn{2}{c}{{\bfseries readability\,$\uparrow$}} & \multicolumn{2}{c}{{\bfseries relevance\,$\uparrow$}} & \multicolumn{2}{c}{{\bfseries toxicity\,$\downarrow$}} \\ 
\cmidrule(lr){2-3} \cmidrule(lr){4-5} \cmidrule(lr){6-7} \cmidrule(lr){8-9} \cmidrule(lr){10-11} \cmidrule(lr){12-13} \cmidrule(lr){14-15}
term & est. & p & est. & p & est. & p & est. & p & est. & p & est. & p & est. & p \\ 
\midrule\addlinespace[2.5pt]
(Int.) & {\textcolor[HTML]{B3B3B3}{+1.294}} & {\textcolor[HTML]{B3B3B3}{<.001}} & {\textcolor[HTML]{B3B3B3}{+0.390}} & {\textcolor[HTML]{B3B3B3}{.002}} & {\textcolor[HTML]{B3B3B3}{-1.745}} & {\textcolor[HTML]{B3B3B3}{<.001}} & {\textcolor[HTML]{B3B3B3}{-0.061}} & {\textcolor[HTML]{B3B3B3}{<.001}} & {\textcolor[HTML]{B3B3B3}{+1.734}} & {\textcolor[HTML]{B3B3B3}{<.001}} & {\textcolor[HTML]{B3B3B3}{-2.122}} & {\textcolor[HTML]{B3B3B3}{<.001}} & {\textcolor[HTML]{B3B3B3}{-1.996}} & {\textcolor[HTML]{B3B3B3}{<.001}} \\ 
\Ba & {\cellcolor[HTML]{543005}{\textcolor[HTML]{FFFFFF}{-2.035}}} & {\textcolor[HTML]{000000}{<.001}} & {\textcolor[HTML]{B3B3B3}{-0.230}} & {\textcolor[HTML]{B3B3B3}{.114}} & {\textcolor[HTML]{B3B3B3}{-0.108}} & {\textcolor[HTML]{B3B3B3}{.080}} & {\cellcolor[HTML]{003C30}{\textcolor[HTML]{FFFFFF}{+0.145}}} & {\textcolor[HTML]{000000}{<.001}} & {\cellcolor[HTML]{543005}{\textcolor[HTML]{FFFFFF}{-1.433}}} & {\textcolor[HTML]{000000}{<.001}} & {\textcolor[HTML]{B3B3B3}{+0.097}} & {\textcolor[HTML]{B3B3B3}{.137}} & {\cellcolor[HTML]{8C510A}{\textcolor[HTML]{FFFFFF}{-0.825}}} & {\textcolor[HTML]{000000}{<.001}} \\ 
\Redd & {\cellcolor[HTML]{E9F2F1}{\textcolor[HTML]{000000}{+0.105}}} & {\textcolor[HTML]{000000}{<.001}} & {\cellcolor[HTML]{2E8E86}{\textcolor[HTML]{FFFFFF}{+0.570}}} & {\textcolor[HTML]{000000}{<.001}} & {\textcolor[HTML]{B3B3B3}{-0.044}} & {\textcolor[HTML]{B3B3B3}{.240}} & {\cellcolor[HTML]{86CFC4}{\textcolor[HTML]{000000}{+0.056}}} & {\textcolor[HTML]{000000}{<.001}} & {\cellcolor[HTML]{52AAA0}{\textcolor[HTML]{FFFFFF}{+0.758}}} & {\textcolor[HTML]{000000}{<.001}} & {\cellcolor[HTML]{003C30}{\textcolor[HTML]{FFFFFF}{+0.502}}} & {\textcolor[HTML]{000000}{<.001}} & {\cellcolor[HTML]{49A49B}{\textcolor[HTML]{FFFFFF}{+0.569}}} & {\textcolor[HTML]{000000}{<.001}} \\ 
\Mu & {\textcolor[HTML]{B3B3B3}{-0.012}} & {\textcolor[HTML]{B3B3B3}{.474}} & {\cellcolor[HTML]{389991}{\textcolor[HTML]{FFFFFF}{+0.533}}} & {\textcolor[HTML]{000000}{<.001}} & {\textcolor[HTML]{B3B3B3}{-0.024}} & {\textcolor[HTML]{B3B3B3}{.603}} & {\textcolor[HTML]{B3B3B3}{+0.008}} & {\textcolor[HTML]{B3B3B3}{.303}} & {\cellcolor[HTML]{F6E8C5}{\textcolor[HTML]{000000}{-0.277}}} & {\textcolor[HTML]{000000}{.001}} & {\cellcolor[HTML]{A1DAD1}{\textcolor[HTML]{000000}{+0.156}}} & {\textcolor[HTML]{000000}{.003}} & {\textcolor[HTML]{B3B3B3}{+0.015}} & {\textcolor[HTML]{B3B3B3}{.834}} \\ 
\Hs & {\cellcolor[HTML]{E1F0EE}{\textcolor[HTML]{000000}{+0.181}}} & {\textcolor[HTML]{000000}{<.001}} & {\textcolor[HTML]{B3B3B3}{-0.160}} & {\textcolor[HTML]{B3B3B3}{.102}} & {\cellcolor[HTML]{683C07}{\textcolor[HTML]{FFFFFF}{-0.237}}} & {\textcolor[HTML]{000000}{<.001}} & {\cellcolor[HTML]{B47726}{\textcolor[HTML]{FFFFFF}{-0.093}}} & {\textcolor[HTML]{000000}{<.001}} & {\cellcolor[HTML]{DDBC76}{\textcolor[HTML]{000000}{-0.598}}} & {\textcolor[HTML]{000000}{<.001}} & {\cellcolor[HTML]{B17324}{\textcolor[HTML]{FFFFFF}{-0.329}}} & {\textcolor[HTML]{000000}{<.001}} & {\cellcolor[HTML]{003C30}{\textcolor[HTML]{FFFFFF}{+1.030}}} & {\textcolor[HTML]{000000}{<.001}} \\ 
\Hist & {\cellcolor[HTML]{C0E7E1}{\textcolor[HTML]{000000}{+0.448}}} & {\textcolor[HTML]{000000}{<.001}} & {\cellcolor[HTML]{003C30}{\textcolor[HTML]{FFFFFF}{+0.899}}} & {\textcolor[HTML]{000000}{<.001}} & {\cellcolor[HTML]{9C6016}{\textcolor[HTML]{FFFFFF}{-0.188}}} & {\textcolor[HTML]{000000}{.002}} & {\cellcolor[HTML]{F6EBCD}{\textcolor[HTML]{000000}{-0.023}}} & {\textcolor[HTML]{000000}{.029}} & {\textcolor[HTML]{B3B3B3}{+0.014}} & {\textcolor[HTML]{B3B3B3}{.920}} & {\textcolor[HTML]{B3B3B3}{+0.072}} & {\textcolor[HTML]{B3B3B3}{.270}} & {\textcolor[HTML]{B3B3B3}{+0.104}} & {\textcolor[HTML]{B3B3B3}{.327}} \\ 
\Prev & {\cellcolor[HTML]{AEE0D8}{\textcolor[HTML]{000000}{+0.554}}} & {\textcolor[HTML]{000000}{<.001}} & {\cellcolor[HTML]{2B8C84}{\textcolor[HTML]{FFFFFF}{+0.579}}} & {\textcolor[HTML]{000000}{<.001}} & {\textcolor[HTML]{B3B3B3}{+0.012}} & {\textcolor[HTML]{B3B3B3}{.820}} & {\cellcolor[HTML]{69BBB0}{\textcolor[HTML]{000000}{+0.068}}} & {\textcolor[HTML]{000000}{<.001}} & {\textcolor[HTML]{B3B3B3}{-0.035}} & {\textcolor[HTML]{B3B3B3}{.750}} & {\cellcolor[HTML]{A5DCD3}{\textcolor[HTML]{000000}{+0.150}}} & {\textcolor[HTML]{000000}{.005}} & {\cellcolor[HTML]{E2C786}{\textcolor[HTML]{000000}{-0.386}}} & {\textcolor[HTML]{000000}{<.001}} \\ 
\Summ & {\cellcolor[HTML]{D1ECE8}{\textcolor[HTML]{000000}{+0.322}}} & {\textcolor[HTML]{000000}{<.001}} & {\cellcolor[HTML]{45A198}{\textcolor[HTML]{FFFFFF}{+0.505}}} & {\textcolor[HTML]{000000}{.001}} & {\textcolor[HTML]{B3B3B3}{+0.011}} & {\textcolor[HTML]{B3B3B3}{.859}} & {\textcolor[HTML]{B3B3B3}{+0.014}} & {\textcolor[HTML]{B3B3B3}{.195}} & {\textcolor[HTML]{B3B3B3}{+0.030}} & {\textcolor[HTML]{B3B3B3}{.813}} & {\cellcolor[HTML]{75C5B9}{\textcolor[HTML]{000000}{+0.216}}} & {\textcolor[HTML]{000000}{<.001}} & {\textcolor[HTML]{B3B3B3}{-0.037}} & {\textcolor[HTML]{B3B3B3}{.729}} \\ 
\Ba:\Hist & {\cellcolor[HTML]{A9DDD5}{\textcolor[HTML]{000000}{+0.583}}} & {\textcolor[HTML]{000000}{<.001}} & {\cellcolor[HTML]{BC7E2B}{\textcolor[HTML]{FFFFFF}{-0.551}}} & {\textcolor[HTML]{000000}{.002}} & {\cellcolor[HTML]{003C30}{\textcolor[HTML]{FFFFFF}{+0.256}}} & {\textcolor[HTML]{000000}{<.001}} & {\cellcolor[HTML]{68BBB0}{\textcolor[HTML]{000000}{+0.068}}} & {\textcolor[HTML]{000000}{<.001}} & {\textcolor[HTML]{B3B3B3}{-0.132}} & {\textcolor[HTML]{B3B3B3}{.362}} & {\textcolor[HTML]{B3B3B3}{+0.118}} & {\textcolor[HTML]{B3B3B3}{.116}} & {\textcolor[HTML]{B3B3B3}{+0.027}} & {\textcolor[HTML]{B3B3B3}{.865}} \\ 
\Ba:\Prev & {\cellcolor[HTML]{E5F1EF}{\textcolor[HTML]{000000}{+0.145}}} & {\textcolor[HTML]{000000}{<.001}} & {\cellcolor[HTML]{DCBA74}{\textcolor[HTML]{000000}{-0.380}}} & {\textcolor[HTML]{000000}{.010}} & {\textcolor[HTML]{B3B3B3}{+0.078}} & {\textcolor[HTML]{B3B3B3}{.179}} & {\cellcolor[HTML]{D7B068}{\textcolor[HTML]{000000}{-0.066}}} & {\textcolor[HTML]{000000}{<.001}} & {\textcolor[HTML]{B3B3B3}{+0.038}} & {\textcolor[HTML]{B3B3B3}{.746}} & {\textcolor[HTML]{B3B3B3}{-0.114}} & {\textcolor[HTML]{B3B3B3}{.057}} & {\textcolor[HTML]{B3B3B3}{-0.134}} & {\textcolor[HTML]{B3B3B3}{.271}} \\ 
\Ba:\Summ & {\cellcolor[HTML]{59AFA6}{\textcolor[HTML]{FFFFFF}{+1.035}}} & {\textcolor[HTML]{000000}{<.001}} & {\textcolor[HTML]{B3B3B3}{-0.003}} & {\textcolor[HTML]{B3B3B3}{.988}} & {\textcolor[HTML]{B3B3B3}{+0.013}} & {\textcolor[HTML]{B3B3B3}{.860}} & {\cellcolor[HTML]{77C6BA}{\textcolor[HTML]{000000}{+0.062}}} & {\textcolor[HTML]{000000}{<.001}} & {\cellcolor[HTML]{ACDFD7}{\textcolor[HTML]{000000}{+0.400}}} & {\textcolor[HTML]{000000}{.004}} & {\textcolor[HTML]{B3B3B3}{+0.021}} & {\textcolor[HTML]{B3B3B3}{.780}} & {\cellcolor[HTML]{1C7B73}{\textcolor[HTML]{FFFFFF}{+0.735}}} & {\textcolor[HTML]{000000}{<.001}} \\ 
\Redd:\Hs & {\textcolor[HTML]{B3B3B3}{-0.007}} & {\textcolor[HTML]{B3B3B3}{.594}} & {\textcolor[HTML]{B3B3B3}{+0.161}} & {\textcolor[HTML]{B3B3B3}{.087}} & {\textcolor[HTML]{B3B3B3}{+0.012}} & {\textcolor[HTML]{B3B3B3}{.713}} & {\cellcolor[HTML]{F0DCAE}{\textcolor[HTML]{000000}{-0.038}}} & {\textcolor[HTML]{000000}{<.001}} & {\cellcolor[HTML]{F1DFB2}{\textcolor[HTML]{000000}{-0.356}}} & {\textcolor[HTML]{000000}{<.001}} & {\cellcolor[HTML]{EED9A8}{\textcolor[HTML]{000000}{-0.139}}} & {\textcolor[HTML]{000000}{<.001}} & {\cellcolor[HTML]{D7AF66}{\textcolor[HTML]{000000}{-0.472}}} & {\textcolor[HTML]{000000}{<.001}} \\ 
\Redd:\Hist & {\cellcolor[HTML]{EEF3F3}{\textcolor[HTML]{000000}{+0.060}}} & {\textcolor[HTML]{000000}{<.001}} & {\cellcolor[HTML]{C89141}{\textcolor[HTML]{FFFFFF}{-0.495}}} & {\textcolor[HTML]{000000}{<.001}} & {\textcolor[HTML]{B3B3B3}{+0.017}} & {\textcolor[HTML]{B3B3B3}{.687}} & {\textcolor[HTML]{B3B3B3}{-0.001}} & {\textcolor[HTML]{B3B3B3}{.888}} & {\cellcolor[HTML]{F4E5BE}{\textcolor[HTML]{000000}{-0.308}}} & {\textcolor[HTML]{000000}{.001}} & {\cellcolor[HTML]{E9D29B}{\textcolor[HTML]{000000}{-0.158}}} & {\textcolor[HTML]{000000}{<.001}} & {\cellcolor[HTML]{F6E9C7}{\textcolor[HTML]{000000}{-0.190}}} & {\textcolor[HTML]{000000}{.014}} \\ 
\Redd:\Prev & {\cellcolor[HTML]{ECF3F2}{\textcolor[HTML]{000000}{+0.081}}} & {\textcolor[HTML]{000000}{<.001}} & {\textcolor[HTML]{B3B3B3}{-0.089}} & {\textcolor[HTML]{B3B3B3}{.343}} & {\textcolor[HTML]{B3B3B3}{-0.001}} & {\textcolor[HTML]{B3B3B3}{.988}} & {\cellcolor[HTML]{F6EDD6}{\textcolor[HTML]{000000}{-0.018}}} & {\textcolor[HTML]{000000}{.003}} & {\cellcolor[HTML]{F6EAC9}{\textcolor[HTML]{000000}{-0.253}}} & {\textcolor[HTML]{000000}{<.001}} & {\textcolor[HTML]{B3B3B3}{+0.020}} & {\textcolor[HTML]{B3B3B3}{.541}} & {\textcolor[HTML]{B3B3B3}{+0.105}} & {\textcolor[HTML]{B3B3B3}{.095}} \\ 
\Redd:\Summ & {\textcolor[HTML]{B3B3B3}{+0.030}} & {\textcolor[HTML]{B3B3B3}{.071}} & {\cellcolor[HTML]{D5AB61}{\textcolor[HTML]{000000}{-0.422}}} & {\textcolor[HTML]{000000}{<.001}} & {\textcolor[HTML]{B3B3B3}{+0.032}} & {\textcolor[HTML]{B3B3B3}{.443}} & {\textcolor[HTML]{B3B3B3}{-0.010}} & {\textcolor[HTML]{B3B3B3}{.168}} & {\cellcolor[HTML]{F6ECD4}{\textcolor[HTML]{000000}{-0.187}}} & {\textcolor[HTML]{000000}{.034}} & {\cellcolor[HTML]{E2C786}{\textcolor[HTML]{000000}{-0.188}}} & {\textcolor[HTML]{000000}{<.001}} & {\textcolor[HTML]{B3B3B3}{-0.048}} & {\textcolor[HTML]{B3B3B3}{.521}} \\ 
\Mu:\Hist & {\cellcolor[HTML]{F0F4F3}{\textcolor[HTML]{000000}{+0.049}}} & {\textcolor[HTML]{000000}{.022}} & {\cellcolor[HTML]{DFC27E}{\textcolor[HTML]{000000}{-0.357}}} & {\textcolor[HTML]{000000}{.011}} & {\textcolor[HTML]{B3B3B3}{+0.069}} & {\textcolor[HTML]{B3B3B3}{.223}} & {\cellcolor[HTML]{84CEC3}{\textcolor[HTML]{000000}{+0.057}}} & {\textcolor[HTML]{000000}{<.001}} & {\cellcolor[HTML]{93D5CA}{\textcolor[HTML]{000000}{+0.498}}} & {\textcolor[HTML]{000000}{<.001}} & {\textcolor[HTML]{B3B3B3}{-0.053}} & {\textcolor[HTML]{B3B3B3}{.384}} & {\cellcolor[HTML]{D7B168}{\textcolor[HTML]{000000}{-0.466}}} & {\textcolor[HTML]{000000}{<.001}} \\ 
\Mu:\Prev & {\textcolor[HTML]{B3B3B3}{-0.014}} & {\textcolor[HTML]{B3B3B3}{.428}} & {\textcolor[HTML]{B3B3B3}{-0.084}} & {\textcolor[HTML]{B3B3B3}{.463}} & {\textcolor[HTML]{B3B3B3}{+0.009}} & {\textcolor[HTML]{B3B3B3}{.846}} & {\cellcolor[HTML]{ECD6A3}{\textcolor[HTML]{000000}{-0.042}}} & {\textcolor[HTML]{000000}{<.001}} & {\textcolor[HTML]{B3B3B3}{-0.017}} & {\textcolor[HTML]{B3B3B3}{.861}} & {\cellcolor[HTML]{F6E8C4}{\textcolor[HTML]{000000}{-0.099}}} & {\textcolor[HTML]{000000}{.043}} & {\textcolor[HTML]{B3B3B3}{0.000}} & {\textcolor[HTML]{B3B3B3}{.998}} \\ 
\Mu:\Summ & {\cellcolor[HTML]{F0F4F3}{\textcolor[HTML]{000000}{+0.049}}} & {\textcolor[HTML]{000000}{.020}} & {\cellcolor[HTML]{E0C480}{\textcolor[HTML]{000000}{-0.351}}} & {\textcolor[HTML]{000000}{.010}} & {\textcolor[HTML]{B3B3B3}{+0.013}} & {\textcolor[HTML]{B3B3B3}{.820}} & {\cellcolor[HTML]{BDE6E0}{\textcolor[HTML]{000000}{+0.033}}} & {\textcolor[HTML]{000000}{<.001}} & {\textcolor[HTML]{B3B3B3}{+0.183}} & {\textcolor[HTML]{B3B3B3}{.108}} & {\cellcolor[HTML]{F1DEB1}{\textcolor[HTML]{000000}{-0.126}}} & {\textcolor[HTML]{000000}{.035}} & {\textcolor[HTML]{B3B3B3}{-0.163}} & {\textcolor[HTML]{B3B3B3}{.080}} \\ 
\Hs:\Hist & {\cellcolor[HTML]{F6F2EA}{\textcolor[HTML]{000000}{-0.093}}} & {\textcolor[HTML]{000000}{<.001}} & {\textcolor[HTML]{B3B3B3}{+0.018}} & {\textcolor[HTML]{B3B3B3}{.878}} & {\cellcolor[HTML]{015349}{\textcolor[HTML]{FFFFFF}{+0.227}}} & {\textcolor[HTML]{000000}{<.001}} & {\cellcolor[HTML]{2B8B83}{\textcolor[HTML]{FFFFFF}{+0.094}}} & {\textcolor[HTML]{000000}{<.001}} & {\cellcolor[HTML]{47A39A}{\textcolor[HTML]{FFFFFF}{+0.797}}} & {\textcolor[HTML]{000000}{<.001}} & {\textcolor[HTML]{B3B3B3}{+0.052}} & {\textcolor[HTML]{B3B3B3}{.198}} & {\cellcolor[HTML]{B17424}{\textcolor[HTML]{FFFFFF}{-0.673}}} & {\textcolor[HTML]{000000}{<.001}} \\ 
\Hs:\Prev & {\cellcolor[HTML]{F6EFDE}{\textcolor[HTML]{000000}{-0.184}}} & {\textcolor[HTML]{000000}{<.001}} & {\textcolor[HTML]{B3B3B3}{+0.006}} & {\textcolor[HTML]{B3B3B3}{.949}} & {\textcolor[HTML]{B3B3B3}{+0.021}} & {\textcolor[HTML]{B3B3B3}{.537}} & {\cellcolor[HTML]{F6EEDB}{\textcolor[HTML]{000000}{-0.015}}} & {\textcolor[HTML]{000000}{.012}} & {\cellcolor[HTML]{C7EAE5}{\textcolor[HTML]{000000}{+0.284}}} & {\textcolor[HTML]{000000}{<.001}} & {\textcolor[HTML]{B3B3B3}{-0.018}} & {\textcolor[HTML]{B3B3B3}{.578}} & {\textcolor[HTML]{B3B3B3}{+0.039}} & {\textcolor[HTML]{B3B3B3}{.540}} \\ 
\Hs:\Summ & {\cellcolor[HTML]{F5F3EE}{\textcolor[HTML]{000000}{-0.054}}} & {\textcolor[HTML]{000000}{<.001}} & {\textcolor[HTML]{B3B3B3}{+0.169}} & {\textcolor[HTML]{B3B3B3}{.132}} & {\cellcolor[HTML]{27877F}{\textcolor[HTML]{FFFFFF}{+0.170}}} & {\textcolor[HTML]{000000}{<.001}} & {\cellcolor[HTML]{8BD1C6}{\textcolor[HTML]{000000}{+0.054}}} & {\textcolor[HTML]{000000}{<.001}} & {\cellcolor[HTML]{61B5AB}{\textcolor[HTML]{000000}{+0.699}}} & {\textcolor[HTML]{000000}{<.001}} & {\cellcolor[HTML]{B3E2DB}{\textcolor[HTML]{000000}{+0.129}}} & {\textcolor[HTML]{000000}{.001}} & {\cellcolor[HTML]{A6691C}{\textcolor[HTML]{FFFFFF}{-0.720}}} & {\textcolor[HTML]{000000}{<.001}} \\ 
\Hist:\Prev & {\cellcolor[HTML]{EED9A8}{\textcolor[HTML]{000000}{-0.562}}} & {\textcolor[HTML]{000000}{<.001}} & {\cellcolor[HTML]{E6CC90}{\textcolor[HTML]{000000}{-0.309}}} & {\textcolor[HTML]{000000}{<.001}} & {\textcolor[HTML]{B3B3B3}{-0.021}} & {\textcolor[HTML]{B3B3B3}{.528}} & {\cellcolor[HTML]{DDEFED}{\textcolor[HTML]{000000}{+0.015}}} & {\textcolor[HTML]{000000}{.010}} & {\textcolor[HTML]{B3B3B3}{-0.041}} & {\textcolor[HTML]{B3B3B3}{.561}} & {\textcolor[HTML]{B3B3B3}{+0.001}} & {\textcolor[HTML]{B3B3B3}{.974}} & {\cellcolor[HTML]{5FB3A9}{\textcolor[HTML]{000000}{+0.509}}} & {\textcolor[HTML]{000000}{<.001}} \\ 
\Prev:\Summ & {\cellcolor[HTML]{EEDAAA}{\textcolor[HTML]{000000}{-0.552}}} & {\textcolor[HTML]{000000}{<.001}} & {\cellcolor[HTML]{F0DEB0}{\textcolor[HTML]{000000}{-0.228}}} & {\textcolor[HTML]{000000}{.009}} & {\textcolor[HTML]{B3B3B3}{-0.057}} & {\textcolor[HTML]{B3B3B3}{.088}} & {\textcolor[HTML]{B3B3B3}{-0.009}} & {\textcolor[HTML]{B3B3B3}{.141}} & {\textcolor[HTML]{B3B3B3}{-0.094}} & {\textcolor[HTML]{B3B3B3}{.164}} & {\cellcolor[HTML]{F6EACB}{\textcolor[HTML]{000000}{-0.085}}} & {\textcolor[HTML]{000000}{.012}} & {\cellcolor[HTML]{7BC9BD}{\textcolor[HTML]{000000}{+0.427}}} & {\textcolor[HTML]{000000}{<.001}} \\ 
\bottomrule
\end{tabular*}
\caption{\rev{Overview of the beta regressions by algorithmic indicator based on all single configuration factors and allowed pairwise interactions. Only factor terms with a statistically significant p-value ($p<.05$) are highlighted. For estimated (est.) means, cell color represents the factor's column-wise relative effect (i.e., at the indicator level) and its sign. All beta regressions use the standard \emph{logit} link function and constant precision ($\phi$). Arrows ($\uparrow$ / $\downarrow$) indicate whether higher or lower values respectively are better for each indicator.}}
\label{tab:indicator-factor-betareg}
\end{table}

\subsubsection{Configurations}
We now turn to a fine-grained analysis of the algorithmic evaluation results for each individual configuration. These results are reported in the Appendix Table~\ref{tab:results-indicators-long}, where configurations are grouped into four categories, from top to bottom: those with neither adaptation nor personalization, those with adaptation only, those with personalization only, and those with both. For each evaluation indicator, the best-performing configuration is shown in bold, while the remaining top five are underlined. Although many configurations yield similar scores, each group contains at least some configurations that perform well across selected indicators. Nonetheless, key differences emerge across the groups. Notably, several configurations that incorporate both adaptation and personalization achieve top results across multiple indicators. Representative examples include [\Mu\Redd\Prev\Hist], [\Mu\Hs\Redd\Hist], and [\Mu\Redd\Hist]. Configurations relying exclusively on adaptation also perform strongly in several cases, for instance, [\Mu\Redd], [\Mu\Redd\Prev], and [\Mu\Hs\Redd\Prev] consistently attain competitive results. In contrast, configurations that rely solely on personalization exhibit weaker and less consistent performance. Interestingly, a few configurations that involve neither adaptation nor personalization still yield solid results. The baseline configuration [\Ba], as well as the version fine-tuned solely on \textsc{MultiCONAN} [\Mu], perform comparably to more complex strategies in several indicators. This underscores that model quality is not determined purely by the number of added factors, but also by how well these are integrated. In sum, while overall performance differences across configurations tend to be modest, the results suggest that applying adaptation---or adaptation in combination with personalization—tends to improve the quality of generated counterspeech more consistently than using personalization alone or applying no contextualization at all. \rev{In the Appendix Section \ref{sub:qwen}, we present a comparative analysis with the counterspeech generated by another model, Qwen3-8B.}

\subsubsection{Extended evaluation of algorithmic indicators} 
\begin{figure*}[t] 
\centering
    \includegraphics[width=0.55\columnwidth]{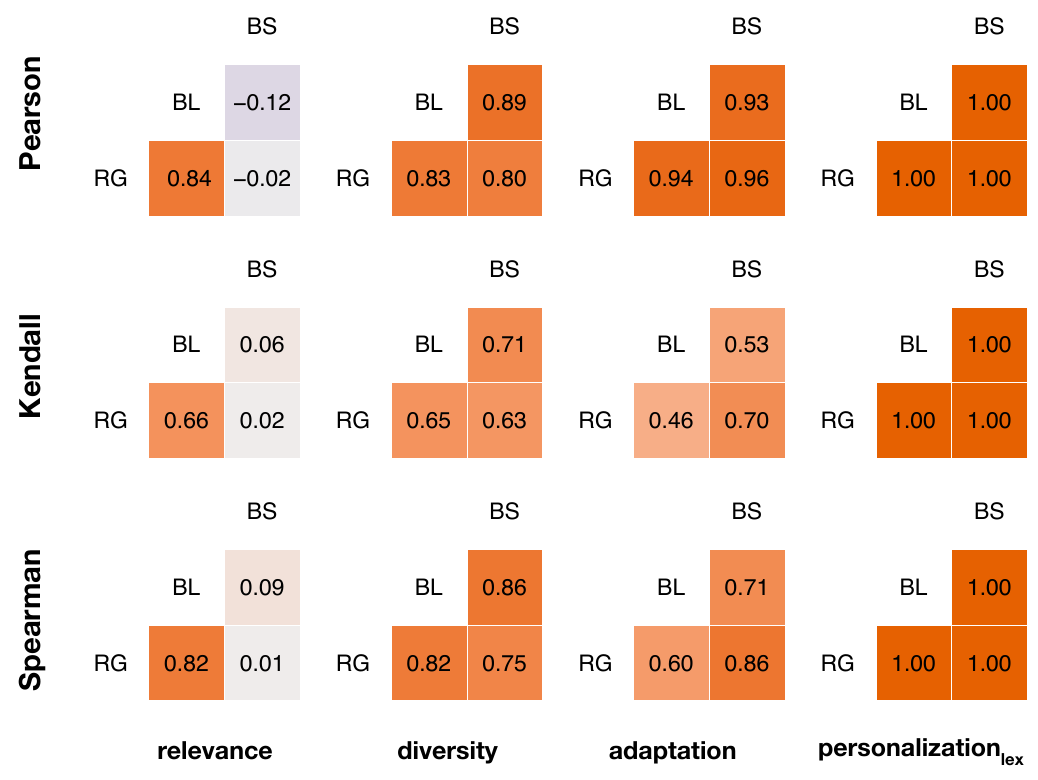}
    \caption{Linear and rank correlations (\textit{y} axis) between four algorithmic indicators (\textit{x} axis) when computed with different text similarity metrics: ROUGE (RG), BLEU (BL), and BERTScore (BS).}
    \label{fig:base-score-corplot}
\end{figure*}

Several algorithmic indicators defined in Section~\ref{sec:method-evaluation-algo} rely on ROUGE to compute text similarities. However, multiple such metrics exist in the literature. To assess the robustness of our findings, we re-implemented all ROUGE-based indicators using BLEU and BERTScore, and compared the results across the three implementations. Figure~\ref{fig:base-score-corplot} reports linear and rank correlations for the four indicators originally based on ROUGE. Overall, we observe strong agreement between implementations. We measured perfect consistency for \textit{lexical personalization} and high correlations for both \textit{adaptation} and \textit{diversity}. In contrast, measuring \textit{relevance} with BERTScore yields results largely uncorrelated with those obtained via ROUGE or BLEU, which remain closely aligned. These results indicate a relatively strong invariance of the considered indicators with respect to the underlying text similarity metric used. 

\subsubsection{Configuration selection} Next, we select a subset of configurations from Table~\ref{tab:results-indicators-long} to further evaluate manually. The selection is carried out via the methodology described in Section~\ref{sec:method-evaluation-selection}. Specifically, we identify the best- and worst-performing configurations within each of the three main strategies: adaptation-only, personalization-only, and the combined approach. These six configurations are highlighted in Table~\ref{tab:results-indicators-long}. In addition to these, we also include the baseline configuration [\Ba] to provide a meaningful reference point for comparison.

\subsection{Human evaluation}
\label{sec:crowdsourcing-results}
We recruited $N=2,444$ and $N=2,353$ participants on Amazon Mechanical Turk for the \textit{non-contextual} and \textit{contextual} between-subjects conditions, respectively. These numbers exclude participants whose responses were rejected due to excessively fast completion times, fewer than $<100\%$ correct answers to control questions, or equal responses across all items. We report no deviations from our pre-registered protocol. \rev{We further assessed the reliability and similarity of crowdworkers' judgments by computing Krippendorff's $\alpha$, pairwise Spearman correlation, pairwise agreement, and normalized match distance, reported in the Appendix Table \ref{tab:kripp}. Results show very low $\alpha$ and correlation values, indicating limited consistency in annotators' relative judgments, but moderate pairwise agreement and low normalized match distance, reflecting that ratings are often concentrated in a narrow range around the middle-high values of the Likert scale.}

\subsubsection{Non-contextual experiment} 
\begin{figure*}[t]
    \begin{subfigure}[t]{1\textwidth}\centering
        \includegraphics[width=1\textwidth]{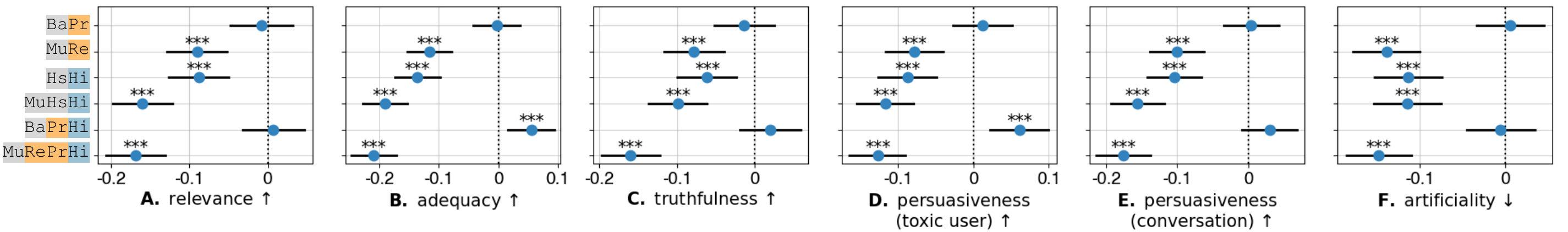}\caption{\textit{Non-contextual} condition.}
        \label{fig:no-context-all}
    \end{subfigure}
\begin{subfigure}[t]{1\textwidth}\centering
        \includegraphics[width=1\textwidth]{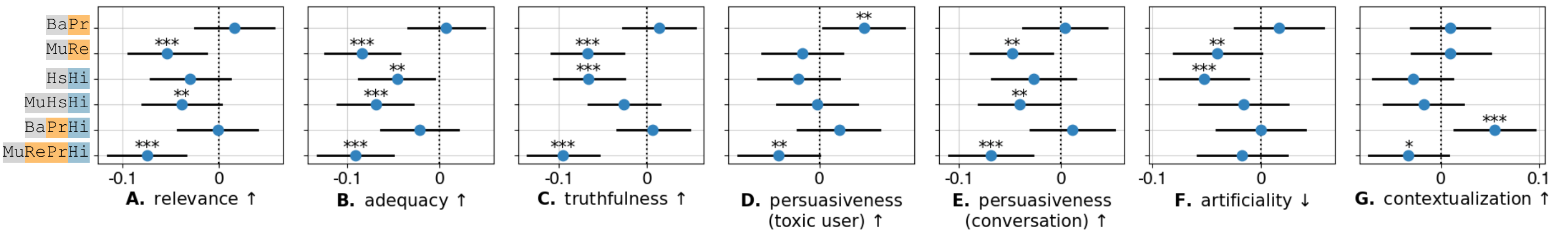}\caption{\textit{Contextual} condition.}
        \label{fig:context-social}
    \end{subfigure}
\caption{Human evaluation results in terms of effect sizes (blue dots) and confidence intervals (black bars) for the scores assigned to various configurations compared to the baseline. Statistical significance levels are indicated as follows: ***: $p < 0.01$, **: $p < 0.05$, *: $p < 0.1$.}
\label{fig:human-results}
\end{figure*}

Participants in the \textit{non-contextual} condition evaluated pairs consisting only of a toxic message and a counterspeech response. A Friedman test reveals statistically significant differences across the evaluated configurations. Figure~\ref{fig:no-context-all} reports the effect sizes, confidence intervals, and statistical significance for each configuration compared to the baseline [\Ba].
The results show two distinct groups of performance. Configurations such as [\Mu\Redd], [\Hs\Hist], [\Mu\Hs\Hist], and [\Mu\Redd\Prev\Hist] consistently perform worse than the baseline across all evaluated aspects, except for \textit{artificiality}, where they are perceived as more human-like. \rev{This pattern is consistent with recent evidence that AI-generated counterspeech can remain distinguishable from human-written counterspeech in terms of linguistic characteristics, politeness, and specificity~\cite{song2025assessing}.} These results suggest that while these models generate responses that seem less machine-generated, they nonetheless produce counterspeech perceived as less relevant, adequate, truthful, or persuasive than that of the baseline. In contrast, configurations [\Ba\Prev] and [\Ba\Prev\Hist] match or exceed the baseline in some dimensions, with statistical significance. Notably, [\Ba\Prev\Hist] performs significantly better in terms of \textit{adequacy} and its perceived capacity to \textit{persuade} the author of the toxic message. It also scores higher (though not significantly) in \textit{relevance}, \textit{truthfulness}, and the perceived ability to \textit{persuade} bystanders. Configuration [\Ba\Prev] performs similarly to the baseline, but yields slightly higher scores in both \textit{persuasiveness} questions. \rev{We emphasize that these results concern perceived persuasiveness: participants judged the likely persuasive effect of each response on another user or on the broader conversation, rather than reporting their own attitude change. Therefore, the findings should be interpreted as evidence of perceived persuasive potential, not as direct evidence of actual persuasion.}

Overall, results from the \textit{non-contextual} within-subjects condition indicate that [\Ba\Prev\Hist] improves upon the baseline in terms of perceived \textit{adequacy} and \textit{persuasiveness}, supporting the value of contextualized counterspeech. Interestingly, the best-performing configurations in this experiment—[\Ba\Prev] and [\Ba\Prev\Hist]—were among the worst ranked by the algorithmic indicators in Table~\ref{tab:results-indicators-long}, suggesting a potential misalignment between automated and human evaluations.

\subsubsection{Contextual experiment} 
Participants in the \textit{contextual} condition received additional information alongside the toxic message and counterspeech, including the subreddit name, the previous message in the thread, and a user summary obtained as described in Section~\ref{sec:method-generation-personalization}. Once again, a Friedman test reveals statistically significant differences across configurations. Detailed results, shown in Figure~\ref{fig:context-social}, largely mirror those from the \textit{non-contextual} experiment.
Configurations [\Ba\Prev] and [\Ba\Prev\Hist] consistently achieve the highest scores. In particular, [\Ba\Prev] yields a statistically significant improvement over the baseline in its ability to \textit{persuade} the author of the toxic message. While most other improvements are not statistically significant, these configurations clearly outperform alternatives such as [\Mu\Redd], [\Hs\Hist], [\Mu\Hs\Hist], and [\Mu\Redd\Prev\Hist], which again show significantly worse performance than the baseline, except for marginal improvements in \textit{artificiality}. This experiment also included an additional item with respect to the \textit{non-contextual} one, evaluating the perceived \textit{contextualization} of each response. Figure~\ref{fig:context-social}\textit{G} shows that [\Ba\Prev\Hist] scores markedly better than the baseline, whereas [\Mu\Redd\Prev\Hist] performs worse, reinforcing previous findings about the benefits of the [\Ba\Prev\Hist] configuration. 

\begin{figure*}[t]
\centering
    \includegraphics[width=1\textwidth]{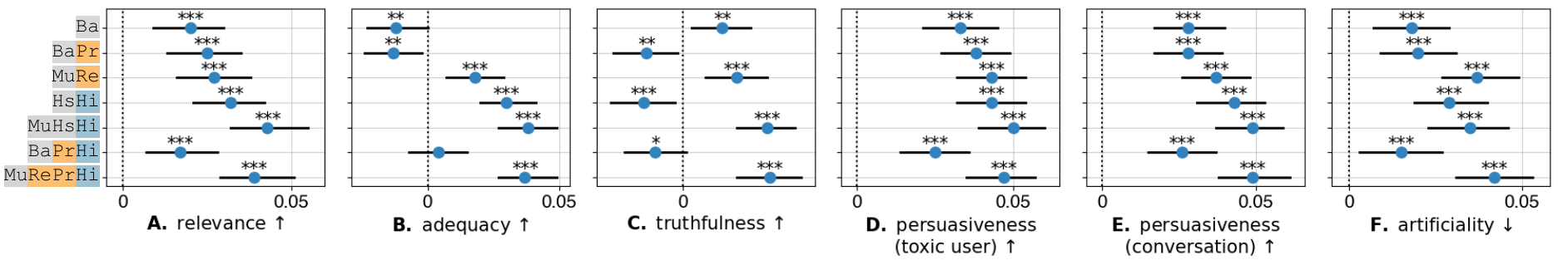}
    \caption{Differences in human evaluation results between the \textit{contextual} and \textit{non-contextual} conditions. Statistical significance: ***: $p < 0.01$, **: $p < 0.05$, *: $p < 0.1$.}
    \label{fig:between-all}
\end{figure*}

Our study design allows for direct comparisons between the \textit{contextual} and \textit{non-contextual} experiments. Figure~\ref{fig:between-all} shows that all configurations—including the baseline—achieve higher ratings in the contextual condition across all dimensions except \textit{adequacy} and \textit{truthfulness}. However, the magnitude of improvement varies. The baseline and configurations that already performed well, such as [\Ba\Prev] and [\Ba\Prev\Hist], show relatively modest gains. In contrast, weaker configurations benefit more substantially from contextual information. 
\rev{These differences may be partly attributable to the evaluation setting itself: providing raters with conversation- and user-level context may help them interpret both the toxic message and the counterspeech response more coherently, leading to generally higher judgments. In addition, the contextual condition included an explicit question about whether the response was personalized rather than generic, which may have made the study hypothesis more salient to participants and potentially influenced their other ratings.}

\begin{figure}[t]
  \centering
  \begin{minipage}[t]{0.52\linewidth}
    \centering
    \includegraphics[width=\linewidth]{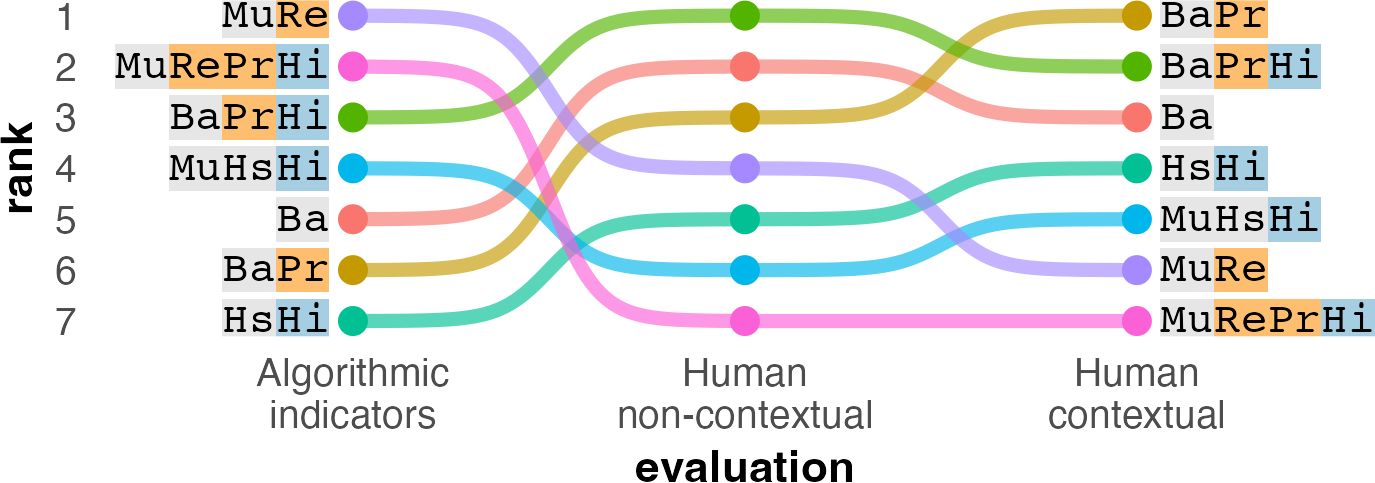}
    \caption{Aggregated rankings of the selected configurations, based on algorithmic and human evaluations. Rankings based on algorithmic indicators are negatively correlated with human evaluations, highlighting a marked mismatch between the two.}
    \label{fig:evaluation-rankings}
  \end{minipage}\hfill
  \begin{minipage}[t]{0.44\linewidth}
    \centering
    \includegraphics[width=0.73\linewidth]{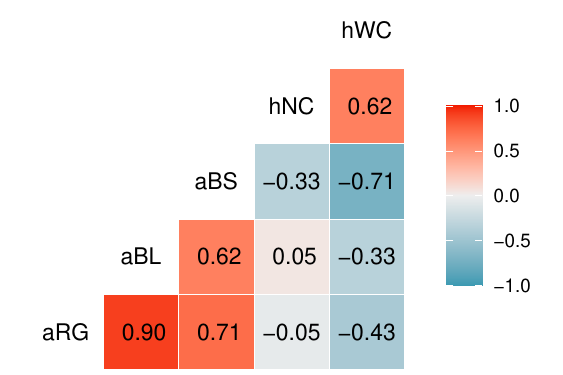}
    \caption{Rank correlation coefficients for the selected model configuration rankings between algorithmic indicators based on ROUGE (aRG), BLEU (aBL) and BERTScore (aBS), and human ones without (hNC) and with context (hWC).}
    \label{fig:selected-model-rankings-corplot}
  \end{minipage}
\end{figure}

\subsubsection{Comparing algorithmic and human evaluations}
We compare the performance of the selected configurations across algorithmic and human evaluations. For each evaluation method (i.e. ROUGE-based quantitative indicators, human assessments with and without context) Figure~\ref{fig:evaluation-rankings} presents the aggregated rankings of configurations across all considered aspects, with the best-performing configurations at the top. Rankings are aggregated using the method described in Section~\ref{sec:method-evaluation-selection}. As shown, the rankings from the two human evaluations are broadly consistent with each other, yielding a Kendall rank correlation $\tau=0.62$. In contrast, the ranking derived from the quantitative indicators diverges markedly from both human evaluations, with negative correlations $\tau=-0.05$ and $\tau=-0.43$ relative to the \textit{non-contextual} and \textit{contextual} experiments, respectively. Figure~\ref{fig:selected-model-rankings-corplot} extends this analysis by incorporating algorithmic indicators computed with BLEU and BERTScore, in addition to those based on ROUGE. The figure reports Kendall rank correlations between every evaluation method. Consistent with Figure~\ref{fig:base-score-corplot}, the different implementations of the algorithmic indicators remain strongly correlated with one another, and the two human-based evaluations also show strong internal agreement. However, correlations between algorithmic and human evaluations are either negligible or, in most cases, moderately to strongly negative. Notably, BERTScore-based evaluations exhibit a strong negative correlation with human judgments in the \textit{contextual} experiment, with $\tau=-0.71$. 
\rev{Our findings provide domain-specific evidence that the well-known evaluation problem of automatic metrics in NLP generation also arises in contextualized counterspeech generation~\cite{zubiaga2024llm}, where key properties such as adequacy, perceived persuasiveness, contextualization, and artificiality are highly subjective and difficult to capture with surface-level similarity measures.}
At the same time, the existence of strong---albeit negative---correlations opens up intriguing opportunities for developing predictive models, for instance via classification or regression, to better approximate human judgments of counterspeech effectiveness, an area that has received little attention so far~\cite{bozdag2026persuade}.

\subsubsection{Impact of toxicity types and emotional profiles on counterspeech persuasiveness} Here we refine the evaluation of the generated counterspeech messages by investigating persuasiveness in relation to the type of toxic speech that the counterspeech corrects, and their emotional profile. 

\begin{figure*}[t]
    \centering
    \begin{subfigure}[t]{0.5\textwidth}\centering
        \includegraphics[width=\textwidth]{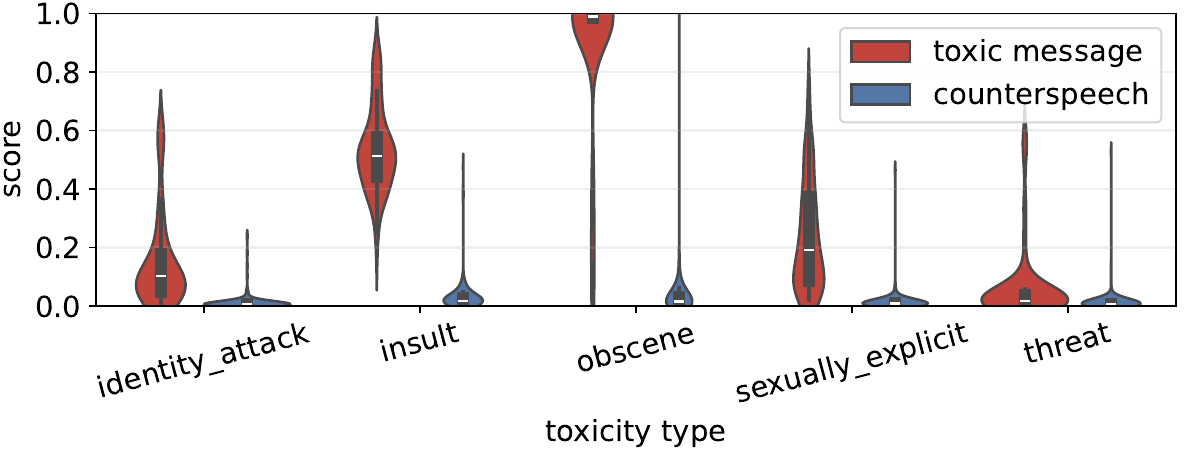}\end{subfigure}\hfill \begin{subfigure}[t]{0.5\textwidth}\centering
        \includegraphics[width=0.5\textwidth]{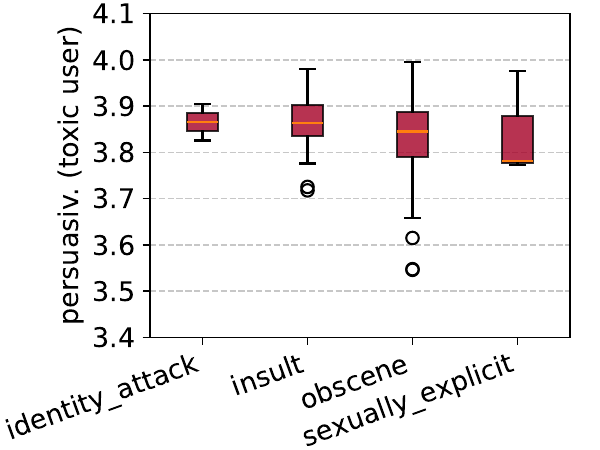}\includegraphics[width=0.5\textwidth]{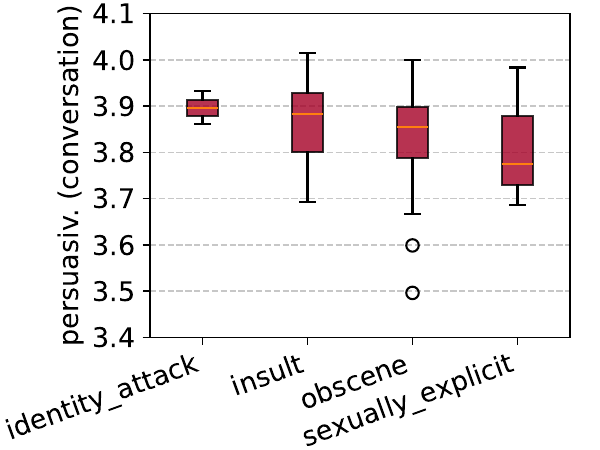}\end{subfigure}
\caption{Distribution of \textbf{toxicity} subtypes in toxic messages and the generated counterspeech (left-side panel) and persuasiveness of the counterspeech with respect to the subtypes of toxic speech in the toxic message (right-side panels).}
\label{fig:dist-pers-toxicity}
\end{figure*}

\begin{figure*}[t]
    \centering
    \begin{subfigure}[t]{0.49\textwidth}\centering
        \includegraphics[width=\textwidth]{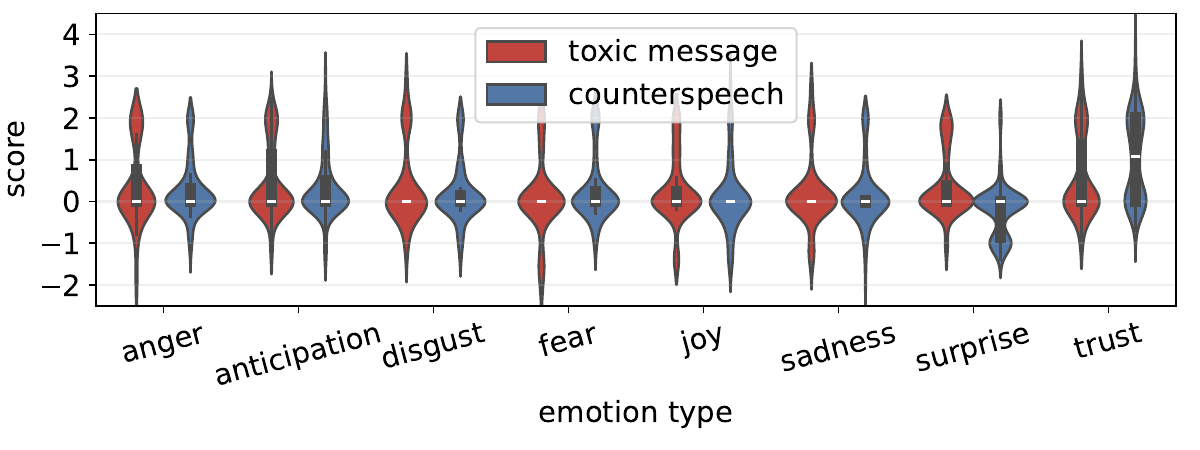}\end{subfigure}\hfill \begin{subfigure}[t]{0.49\textwidth}\centering
        \includegraphics[width=0.5\textwidth]{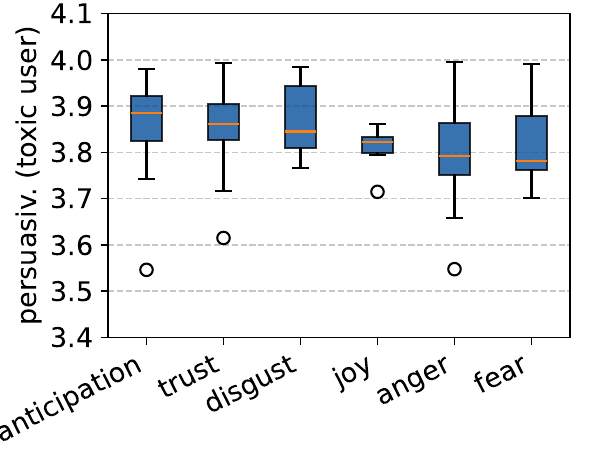}\includegraphics[width=0.5\textwidth]{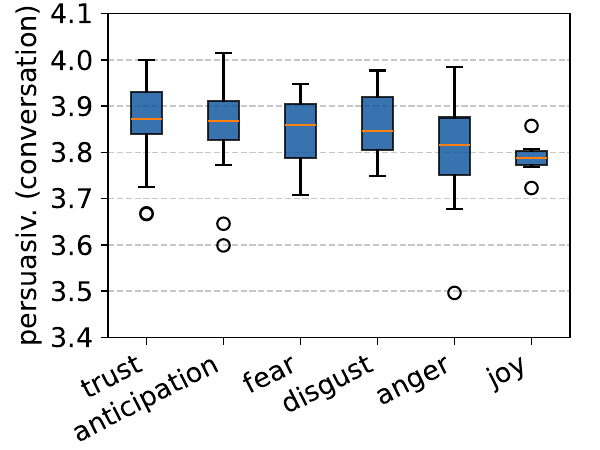}\end{subfigure}\caption{Distribution of \textbf{emotions} subtypes in toxic messages and the generated counterspeech (left-side panel) and persuasiveness of the counterspeech with respect to the dominant emotion of the counterspeech message (right-side panels).}
\label{fig:dist-pers-emotions}
\end{figure*}

\paragraph{Types of toxicity} In addition to the overall toxicity score used so far, Perspective API also provides scores for the following subtypes of toxic speech: \textit{identity attack}, \textit{insult}, \textit{obscene}, \textit{sexually explicit}, and \textit{threat}. 
\rev{The left-side panel of Figure~\ref{fig:dist-pers-toxicity} provides a quantitative characterization of the toxicity profile of the 128 toxic messages in our dataset by reporting their subtype scores. The same panel also reports the corresponding scores for the generated counterspeech. As expected, counterspeech messages exhibit overall low toxicity scores, while the original toxic messages display substantially higher values across several toxicity dimensions. Among the toxic messages, \textit{obscene} content is the most prevalent subtype, followed by \textit{insult} and \textit{sexually explicit} content. This distribution clarifies the scope of our benchmark and indicates that the evaluation is mostly shaped by obscene and insulting toxic language, whereas other toxicity subtypes, such as \textit{threat}, are less represented.}

Based on these scores, we then identified which toxicity subtypes are more effectively mitigated by counterspeech. For this purpose, we define the dominant toxicity subtype of a message as the one with the highest score among those assigned to that message by Perspective API. The right-side panels of Figure~\ref{fig:dist-pers-toxicity} show the persuasiveness scores obtained by the generated counterspeech messages with respect to the subtypes of toxic speech of the corresponding toxic message.\footnote{\textit{Threat} is excluded from this analysis due to its limited presence in our data.} As shown, the counterspeech that we generated appears most effective when responding to \textit{identity attacks}. In contrast, responses to \textit{sexually explicit} content achieved the lowest median ratings. While this pattern suggests that certain types of toxicity may be more amenable to persuasive intervention, the differences are not statistically significant, likely due to the relatively small size of our sample, which amounts to 140 counterspeech instances.

\paragraph{Emotions} We obtain the emotional profile of both toxic and counterspeech messages with EmoAtlas,\footnote{\url{https://github.com/massimostel/emoatlas}} a framework to detect the presence of eight core emotions: \textit{anticipation}, \textit{anger}, \textit{disgust}, \textit{fear}, \textit{joy}, \textit{sadness}, \textit{surprise}, and \textit{trust}. EmoAtlas returns Z-scores representing the intensity of each emotion relative to its training corpus. The left-side panel of Figure~\ref{fig:dist-pers-emotions} shows the distributions of these emotions. We note that both toxic messages and counterspeech overall present more positive scores than negative ones, indicating that both types of messages generally exhibit stronger emotions than the EmoAtlas baseline. Furthermore, the emotions distributions of toxic and counterspeech messages are very similar, implying that, overall, the counterspeech mirrored the emotional profile of the comments it addressed. Next, we identified the dominant emotion in each toxic and counterspeech message as the one with the largest Z-score. This allowed to assess possible variations in persuasiveness of the counterspeech with respect to the dominant emotion of the counterspeech itself or the toxic message it responds to. Results shown in Appendix Figure~\ref{fig:comment-persuasion-emotion} reveal that the dominant emotion in a toxic message does not significantly affect the perceived persuasiveness of the counterspeech response. Instead, the right-side panels of Figure~\ref{fig:dist-pers-emotions} show marked differences based on the dominant emotion of the counterspeech message.\footnote{\textit{Sadness} and \textit{surprise} are excluded from this analysis due to their limited presence in our data.} Counterspeech conveying \textit{anticipation} and \textit{trust} exhibit the highest median \textit{persuasiveness} scores for both the user- and conversation-oriented evaluations. In contrast, counterspeech expressing \textit{anger} is consistently and significantly less persuasive than others ($p < 0.05$, Wilcoxon test). This result suggests that crafting counterspeech that conveys positive emotions may be a promising strategy to enhance persuasiveness. Interestingly however, counterspeech expressing \textit{joy} also scores relatively low, especially for \textit{persuasiveness} toward the conversation ($p < 0.01$), possibly indicating that excessive positivity may feel out of context or less credible in confrontational settings.

\begin{figure*}[t]
    \begin{subfigure}{0.315\textwidth}\centering
            \includegraphics[width=\textwidth]{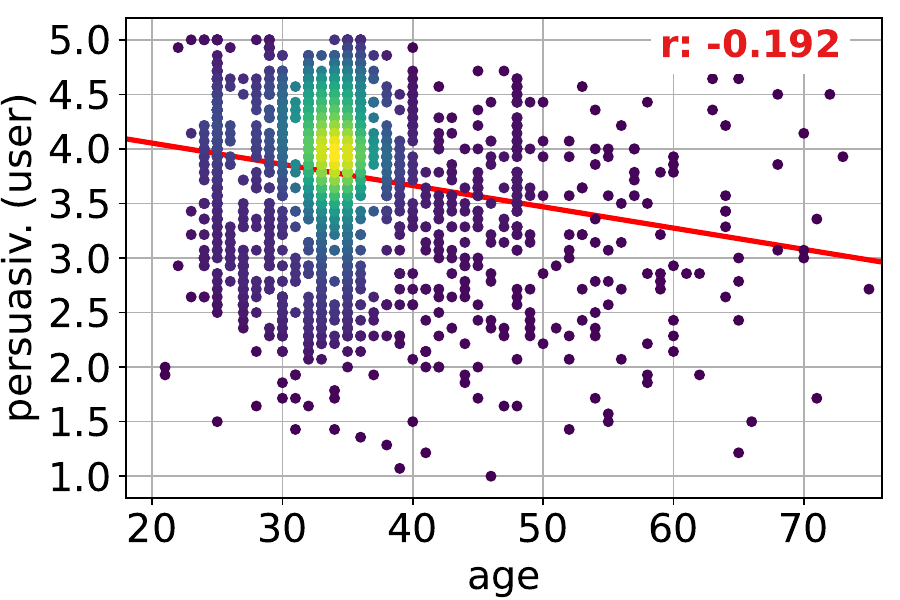}\end{subfigure}\begin{subfigure}{0.315\textwidth}\centering
            \includegraphics[width=\textwidth]{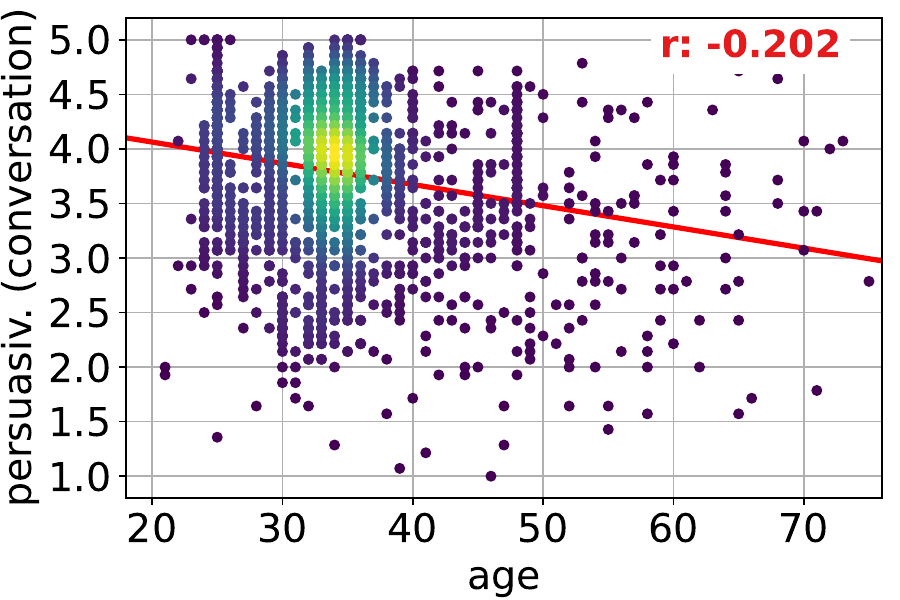}\end{subfigure}\begin{subfigure}{0.315\textwidth}\centering
            \includegraphics[width=\textwidth]{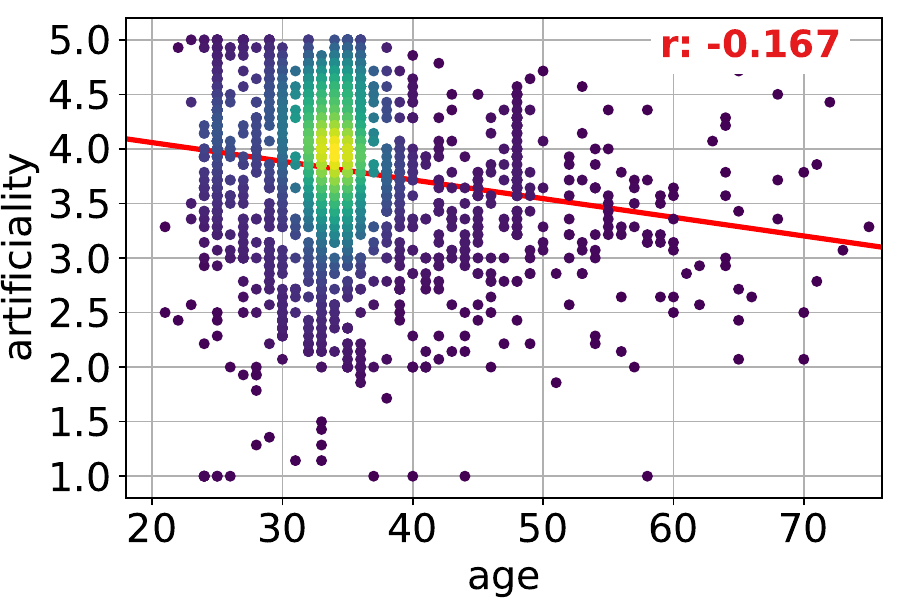}\end{subfigure}\begin{subfigure}[t]{0.03\textwidth}\centering
\includegraphics[width=\textwidth, height=80pt]{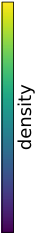}\end{subfigure}\caption{Relationship between respondents' \textit{age} and their judgments: \textit{persuasiveness} with respect to the toxic user (left panel), \textit{persuasiveness} with respect to the conversation (middle panel), and \textit{artificiality} of the counterspeech (right panel). Panels also report Pearson correlations $r$ (statistically significant at $p<0.01$).}
    \label{fig:age-pers}
\end{figure*}

\subsubsection{Age-related differences in perceptions of counterspeech}
Our questionnaire comprised a set of socio-demographic questions,\footnote{The socio-demographic profile of the respondents is shown Appendix Figure~\ref{fig:socio-demo-distribution}.} including the \textit{age} of the participant, as that might influence user evaluations. To investigate its possible impact, we analyzed whether increases in respondents' age correlate with differences in their assessments. We grouped together the \textit{non-contextual} and \textit{contextual} responses and we computed the average score that each participant assigned across the 14 randomly selected counterspeech messages they evaluated. As shown in Figure~\ref{fig:age-pers}, there is a noticeable tendency for older respondents to assign lower scores. This trend is reflected in the moderately negative yet significant ($p < 0.01$) correlations observed between \textit{age} and three key dimensions: \textit{persuasiveness} with respect to the toxic user ($r = -0.192$), \textit{persuasiveness} with respect to the conversation ($r = -0.202$), and \textit{artificiality} ($r = -0.167$). Several factors can explain this phenomenon. Older individuals may be more wary of AI-generated content, perceiving messages as less trustworthy or authentic. They might also be less accustomed to the informal or emotionally charged communication style often used in social media and counterspeech, which could make such messages feel less persuasive or even intrusive. Additionally, generational differences in digital literacy and familiarity with automated systems could influence expectations and evaluations, leading to a more critical stance toward the perceived quality or intent of AI interventions~\cite{zubair2025artificial}.

\subsection{\rev{Failure modes and fine-tuning effects}}
\label{sec:error-analysis}
\rev{The human evaluation identifies which configurations degrade perceived counterspeech quality, but does not explain the source of the degradation. Here we characterize the failure modes of the generated messages through automatic indicators computed over all $6{,}912$ generations. Operationally, we label a generated counterspeech message as:
\begin{itemize}[leftmargin=*]
    \item \emph{Toxic}, if its Perspective API score $>0.5$. This is the same threshold used in Section~\ref{sec:dataset} to select the toxic comments.
    \item \emph{Degenerate}, if it is shorter than ten tokens, truncated, or copies a verbatim $n$-gram ($n \geq 8$) from the toxic message.
    \item \emph{Unframed}, if it contains no marker of a moderation act. That is, no second-person address combined with a directive form, no explicit normative appeal, and no imperative opening.
\end{itemize}
The \emph{unframed} indicator captures a property of register rather than of pragmatic function, and we therefore validated it before use, as per Appendix Section~\ref{sub:unframed-validation}.}

\begin{table}[t]
    \footnotesize
    \centering
    \setlength{\tabcolsep}{6pt}
    \begin{tabular}{lrrrr}
        \toprule
        \textbf{configuration} & \textbf{toxic} $\downarrow$ & \textbf{degenerate} $\downarrow$ & \textbf{unframed} $\downarrow$ & \textbf{length} \\
        \midrule
        \Ba                     & .000 & .008 & .000 & 56 \\
        \Ba\Prev                & .000 & .000 & .000 & 59 \\
        \Ba\Prev\Hist           & .000 & .008 & .000 & 74 \\
        \midrule
        \Mu\Redd                & .133 & .266 & .805 & 19 \\
        \Hs\Hist                & .055 & .273 & .359 & 12 \\
        \Mu\Hs\Hist             & .031 & .195 & .656 & 14 \\
        \Mu\Redd\Prev\Hist      & .086 & .406 & .906 & 17 \\
        \bottomrule
    \end{tabular}
    \caption{\rev{Automatic failure indicators for the seven configurations submitted to human evaluation, computed over all 128 generations of each configuration. Values are proportions. \emph{Length} is the mean generation length in tokens. Lower values are better for all indicators.}}
    \label{tab:failure-rates}
\end{table} 
\rev{\paragraph{Failure rates and relation to human evaluation.} Table~\ref{tab:failure-rates} reports the resulting failure rates for the seven configurations submitted to human evaluation, while the Appendix Table~\ref{tab:counterspeech-examples} illustrates each failure mode with representative outputs. The three configurations based on the non-fine-tuned model never exceed the toxicity threshold and produce no unframed output over the $128$ toxic messages, whereas all fine-tuned configurations produce both. Note that this is a threshold-based count: base configurations do produce some mildly toxic outputs, with maximum toxicity $=0.43$, but none of them reaches the threshold used to select the toxic comments themselves, which is consistent with the mean toxicity scores reported in Table~\ref{tab:results-indicators-long}. In [\Mu\Redd\Prev\Hist], by contrast, $90.6\%$ of the responses contain no moderative framing and $8.6\%$ are themselves toxic. For the fine-tuned configurations, the absence of a moderation act is thus the prevalent outcome rather than an occasional error. Notably, [\Mu\Redd\Prev\Hist] ranks among the best configurations under the algorithmic indicators (Section~\ref{sec:algorithmic-results}), which further illustrates the misalignment between automatic metrics and both human judgment and functional adequacy. The Appendix Table~\ref{tab:counterspeech-examples} additionally documents a failure type that our indicators do not capture, namely factually incorrect statements, which remains a target for future automatic evaluation.} 

\rev{The messages submitted to annotators were the twenty closest to each configuration's centroid in indicator space (Section~\ref{sec:method-evaluation-selection}), which raises the question of whether this procedure excluded the failures reported above. Upon inspection, we verified that it did not exclude the prevalent one: out of the centroid-proximal generations of [\Mu\Redd\Prev\Hist] and [\Mu\Redd], $85\%$ and $70\%$ respectively are unframed, so participants did rate failing outputs. The selected sample is nonetheless cleaner than the non-selected one (pooled unframed rate $0.329$ against $0.401$; stratified permutation test, $p = 0.015$), and none of the $140$ evaluated messages exceeds the toxicity threshold (maximum toxicity $=0.41$), whereas between $3.1\%$ and $13.3\%$ of the full output pools of the fine-tuned configurations do. This shows that toxic generations are extreme in indicator space and were removed by centroid selection. The degradation reported in Section~\ref{sec:crowdsourcing-results} should therefore be read as a conservative estimate, and the most severe failure mode of the pipeline is not observable within the human study, which motivates conducting the present analysis on the full generation pool.}

\rev{\paragraph{Fine-tuning failure.} We consider three explanations for why fine-tuning degrades performance. The first is instruction overload: the model may struggle to integrate multiple concurrent signals~\cite{beck2024sensitivity,giorgi2024human}. Our factorial design tests this directly, since [\Ba\Prev\Hist] and [\Mu\Redd\Prev\Hist] receive the same toxic message, conversational context, and user history, but differ only in their weights. The former is never unframed, whereas the latter is unframed in $90.6\%$ of cases (McNemar's test, $p \approx 3 \times 10^{-31}$; $116$ messages fail only under fine-tuning, and none in the opposite direction). Moreover, within the non-fine-tuned family, adding conversational context and user history never produces unframed outputs ([\Ba], [\Ba\Prev], [\Ba\Hist], and [\Ba\Prev\Hist] are all at $0.000$).\footnote{\rev{The summary factor [\Summ] is an exception, as it produces truncated outputs in over half of the cases when combined with the base model. Since no [\Summ] configuration entered the human evaluation, we do not analyse it further.}} Instruction load is therefore not the operative factor. However, a related source of failure may lie in the model's difficulty in identifying and using the relevant contextual information needed to produce suitable counterspeech~\cite{ricco2026geometric}.}

\rev{The second explanation is toxic priming: fine-tuning on Reddit data, especially when combined with user history, may lead the model to reproduce the target's linguistic behaviour~\cite{cima2025contextualized}. A logistic model of output toxicity over the full design ($N = 6{,}912$) supports this account only in part. Reddit fine-tuning increases toxicity as a main effect (OR $= 4.7$, $p < 0.001$), but its interaction with the comment-history factor is negative and significant ($\beta = -0.80$, $p = 0.001$): toxicity is higher for [\Redd] without user history ($9.9\%$) than with user history ($7.6\%$).\footnote{\rev{Since [\Redd] never occurs without [\Mu] in our design, its coefficient is to be read as the additional effect of Reddit fine-tuning given MultiCONAN fine-tuning.}} Thus, toxicity is attributable to the fine-tuning corpus rather than to conditioning on the moderated user. This is consistent with Appendix Figure~\ref{fig:training-toxicity}: $19.36\%$ of the Reddit comment--reply pairs used for community adaptation exceed the Perspective API toxicity threshold, compared with $4.95\%$ in MultiCONAN and $1.46\%$ in RHSI. Reddit fine-tuning therefore exposes the model to toxic conversational material, plausibly contributing to toxic outputs and to the shift toward the toxic register.}

\rev{The third explanation, which the data support, concerns the register of the generated text. A factorial logistic model over the full design shows that the loss of moderative framing is driven primarily by MultiCONAN fine-tuning (OR $= 7.3$, $p < 0.001$) and, to a lesser extent, by Reddit fine-tuning (OR $= 2.1$, $p < 0.001$), whereas RHSI has no significant effect (OR $= 0.9$, $p = 0.13$). Toxicity, by contrast, is driven by Reddit fine-tuning alone. MultiCONAN contains argumentative counter-narratives that rebut the toxic message without addressing its author, thereby removing the moderative frame ([\Mu] is unframed in $82.8\%$ of cases, against $0.0\%$ for [\Ba]). Reddit comment--reply pairs instead model ordinary conversational participation and introduce both further loss of framing and toxicity ($13.3\%$ toxic for [\Mu\Redd] against $0.8\%$ for [\Mu]). The configuration [\Mu\Redd\Prev\Hist] combines both sources.}

\rev{This register shift is also visible linguistically. Non-fine-tuned configurations produce longer, multi-sentence, and explicitly normative interventions (mean length $72.8$ tokens; $96\%$ with a normative appeal), whereas fine-tuned ones produce shorter responses ($16.8$ tokens; $20\%$). ProfilingUD~\cite{brunato2020profiling} confirms this shift: imperative mood decreases from $32.7$ to $6.1$, subordinate clauses from $8.9$ to $2.2$, and sentences per message from $5.6$ to $1.4$, with all differences significant after FDR correction. Moreover, only the Reddit-fine-tuned configurations move toward the toxic register when compared with the $128$ toxic messages (mean distance $25.4$, against $29.0$ for non-fine-tuned configurations and $28.8$ for fine-tuned ones without Reddit data). Thus, stylistic mirroring is specific to Reddit fine-tuning. This also explains why fine-tuned configurations are perceived as less artificial but judged less adequate and persuasive: their outputs resemble ordinary conversational turns, which lowers artificiality but conflicts with the function of a moderation intervention.}

\rev{\paragraph{Best configurations and implications.} A complementary question is whether the higher ratings obtained by [\Ba\Prev\Hist] reflect contextual fit or, more simply, the fact that without domain fine-tuning the model produces neutral-sounding responses that evaluators find credible. Our data support a qualified version of the latter interpretation. Non-fine-tuned configurations are markedly formulaic: their generations are more similar to one another than those of fine-tuned configurations (mean pairwise similarity $0.383$ against $0.207$), and $90\%$ contain at least one canonical moderation phrase. They are not, however, generic in content, since their lexical overlap with the toxic message is higher than for fine-tuned configurations ($0.092$ against $0.064$), while the [\Mu\Redd] family reaches the highest content specificity ($0.147$) without producing any moderative frame. Participants therefore appear to reward not vagueness, but the recognizable register of a moderation act, despite these configurations receiving the highest artificiality scores. This effect holds at the configuration level: register features correlate with mean adequacy across the seven evaluated configurations ($\rho = 0.93$ for imperative mood, $n = 7$), but not within configurations (all pooled within-configuration correlations $|\rho| < 0.20$). Thus, the register hypothesis accounts for configuration-level rankings, but not for message-level rating variation.}

\rev{Overall, these results indicate that fine-tuning is not neutral with respect to the pragmatic function of the generated text, and that the choice of the fine-tuning corpus determines which aspect of that function is lost. Notably, the best-performing configuration is itself contextualized. What fails is fine-tuning as a vehicle for contextualization, not contextualization as such, suggesting that contextual information is more safely supplied at inference time than encoded in the model weights. Moreover, safety and quality degrade jointly: the toxic outputs of the [\Redd] configurations are the extreme of a distributional shift whose most common outcome is a fluent, non-toxic response that performs no moderation act at all, and a toxicity filter applied downstream would remove the $8$--$13\%$ of harmful outputs while leaving the $80$--$90\%$ that do not function as counterspeech.}

 \section{Discussion and Conclusions}
\label{sec:discussion}
\paragraph{Generation} \rev{Our extensive analysis of adaptation and personalization strategies, evaluated through both algorithmic metrics and large-scale human studies, highlights both the promise and the fragility of generating effective contextualized counterspeech. The human evaluation shows that contextualization is not uniformly beneficial: only a small subset of configurations, especially [\Ba \Prev] and [\Ba \Prev \Hist], matched or improved over the baseline on \textit{adequacy} and perceived \textit{persuasiveness}, whereas several other adaptation and personalization strategies significantly underperformed. This finding suggests that lightweight contextual prompting can improve counterspeech quality when conversational context and user-history information are incorporated in a controlled way. At the same time, the weaker performance of several other configurations indicates that adding more contextual information or applying additional fine-tuning does not necessarily lead to better counterspeech. Current difficulties in consistently generating high-quality contextualized responses may stem from the load placed on LLMs when they must integrate multiple concurrent instructions or layers of information~\cite{giorgi2024human,beck2024sensitivity}. 
These results should be interpreted in light of the ecological-validity gap between crowdsourced perception ratings and actual behavioral change. Our evaluation captures third-party judgments of counterspeech quality, including perceived adequacy and perceived persuasiveness, but it does not measure whether the author of the toxic message would actually change their behavior after receiving the response. This distinction is crucial because LLM-generated contextualized counterspeech can backfire, as highlighted in other works~\cite{bar2024generative}.
Finally, our error analysis shows that some configurations can produce inadequate, incorrect, or even toxic responses. Thus, our results identify contextualized counterspeech as a promising but challenging direction: personalization can be effective in specific settings, but it requires careful design, controlled conditioning, and human-centered evaluation. Future advances in larger and more capable LLMs may reduce some of these limitations, but they will still need to be accompanied by safeguards and rigorous evaluation~\cite{cresci2022personalized}.}

\paragraph{Evaluation} Our findings also carry important implications for how counterspeech systems should be evaluated. While many studies rely on algorithmic metrics to assess the quality of generated counterspeech, our results show that these indicators correlate poorly with human judgments, an observation that aligns with other recent findings~\cite{zubiaga2024llm,hengle2025cseval}. This mismatch suggests that automatic indicators and human evaluators may attend to different, and sometimes orthogonal, qualities of a response. \rev{The low inter-rater reliability further reinforces this point: even when ratings are numerically close, annotators often differ in how they rank or interpret counterspeech quality. This heterogeneity suggests that counterspeech effectiveness may depend not only on the toxic message itself, but also on the expectations, background, and preferences of the people evaluating or receiving the intervention, further motivating personalized and audience-aware approaches.} This discrepancy underscores the importance of developing more comprehensive and nuanced evaluation protocols that combine both algorithmic and human-centered assessments~\cite{bozdag2026persuade}. Relying exclusively on either type may lead to incomplete or misleading conclusions about a system's true effectiveness. Future work should prioritize this endeavor to ensure more accurate assessments of real-world impact. Meanwhile, our findings also show that contextualized, AI-generated counterspeech can be persuasive and impactful according to human evaluation when appropriately adapted and personalized. This suggests a promising direction for scalable, AI-driven interventions aimed at curbing online toxicity. At the same time, this study highlights the limitations of current evaluation methods and points to the need for human-AI collaboration in both the design and assessment of such tools. By combining the scale and adaptability of AI with the nuance of human judgment, future systems could be more effective in promoting healthy online discourse.

\paragraph{Limitations and future work} \rev{While our study provides extensive empirical insights, it is constrained by several limitations. Firstly, we experimented with two large language models and a limited set of adaptation and personalization strategies; alternative architectures or contextualization methods could lead to different results. \rev{In addition, the toxic-message benchmark is numerically limited and restricted to five U.S.-politics-oriented Reddit communities. Its toxicity profile is also skewed toward obscene and insulting content, as shown in Figure~\ref{fig:dist-pers-toxicity}, which may limit generalization to other platforms, cultural contexts, languages, and toxicity types.} Similarly, our algorithmic evaluation is restricted to a limited set of quantitative indicators, and to one generation for each toxic message. \rev{Several of these indicators rely on ROUGE-based lexical similarity to approximate constructs such as relevance, diversity, adaptation, and personalization, raising construct-validity concerns. Moreover, since automatic indicators were also used in the selection pipeline, some configurations or messages preferred by human evaluators may have been excluded. Future work should therefore consider random, stratified, or human-in-the-loop selection strategies and stronger validation of automatic metrics, and should consider multiple generations to capture stochastic variance.} Human evaluations are also subject to crowdsourcing limitations, including sample representativeness, cultural bias, and response variability. \rev{In particular, our persuasiveness scores capture third-party perceptions of persuasive potential rather than actual attitude or behavior change, and may be influenced by the demographic and political composition of the crowdworker sample. Our non-parametric tests also do not jointly model participant- and stimulus-level variability in ordinal ratings; future analyses could complement them with cumulative link mixed models.} These limitations call for further research on contextualized counterspeech using broader models, datasets, and evaluation designs. Future work should also investigate fairness and bias in generated counterspeech, as well as longitudinal or field studies assessing its effects on toxic users, bystanders, and subsequent conversational behavior.}

\section{\rev{Ethics, Risks, and Unintended Harms}}
\label{sec:ethics}
\rev{We investigate AI-generated counterspeech as a tool to support healthier online conversations, but we acknowledge that the same technical components may introduce certain ethical risks and unintended harms. In particular, our models rely on user histories to generate personalized responses, which may raise concerns about behavioral profiling and privacy. However, our goal is limited to capturing topics of discussion and writing style from the target users' prior comments, without inferring sensitive or protected attributes. Moreover, our data collection relies exclusively on publicly available Reddit comments. The proposed approach also has dual-use potential. Although our intended use is moderation support and harm reduction, techniques for contextualized and personalized counterspeech could be repurposed to generate manipulative, deceptive, or politically targeted persuasion at scale~\cite{goldstein2024persuasive,zugecova2024evaluation}. A further risk concerns bias. Since LLMs are known to be bias-prone~\cite{giorgi2024human}, generated counterspeech may reproduce stereotypes, treat communities unevenly, or respond differently depending on the political, cultural, or linguistic characteristics of users and conversations. For these reasons, any real-world deployment should include safeguards such as transparency about AI involvement, human review, limits on user profiling, audits for biased or manipulative outputs, and mechanisms for users and communities to contest or opt out of automated interventions. Furthermore, deployment should be assessed against the current data-access policies and terms of service of the target platform, as well as applicable privacy and data-protection regulations. In light of these risks, we remark that our experiments were pre-registreted and received ethical approval from CNR's IRB (protocol \#0306210). Participants were provided detailed information about the study's purposes, including being informed that participation entailed exposure to toxic comments. All participants gave their informed consent.}
 \section{Acknowledgments}
This work is partially supported by the European Union -- NextGenerationEU within the ERC project DEDUCE (\textit{Data-driven and User-centered Content Moderation}) under grant \#101113826; the PRIN 2022 project PIANO (\textit{Personalized Interventions Against Online Toxicity}) under CUP~B53D23013290006; and the the PNRR MUR project FAIR: \textit{Future AI Research} (PE00000013). Partial support was also received by the MUR in the framework of the FoReLab project (Departments of Excellence) and by the project \textit{``Advancing Italian Language Processing with Small-Scale Training and Preference Modeling''} (IsCb8\_AILP), funded by CINECA under the ISCRA initiative, for the availability of HPC resources and support.

\bibliographystyle{ACM-Reference-Format}
\bibliography{mybib}

\newpage
\begin{appendix}

\section*{Appendix}

\section{\rev{Dataset}}
\rev{For counterspeech generation, we retrieved 128 toxic comments from 49 Reddit threads. Table~\ref{tab:dataset} reports the distribution of these toxic comments across the corresponding subreddits.}

\begin{table}[th]
    \footnotesize
    \centering
    \setlength{\tabcolsep}{3pt}
    \adjustbox{max width=\columnwidth}{
    \begin{tabular}{lccc}
        \toprule
        \textbf{subreddit} & \textbf{threads} & \textbf{toxic comments} \\
        \midrule
        \subr{AOC} & 7 & 8  \\
        \subr{conservatives} & 9 & 15  \\
        \subr{politics} & 8 & 85  \\
        \subr{progressive} & 3 & 4  \\
        \subr{the\_donald} & 8 & 16  \\
        \midrule
        \textit{TOTAL} & 35 & 128 \\
        \bottomrule
    \end{tabular}
    }
    \caption{\rev{Distribution of the toxic comments and relative threads through the five political subreddits analyzed.}}
    \label{tab:dataset}
\end{table}

\begin{figure}[th]
    \centering
    \includegraphics[width=0.4\textwidth]{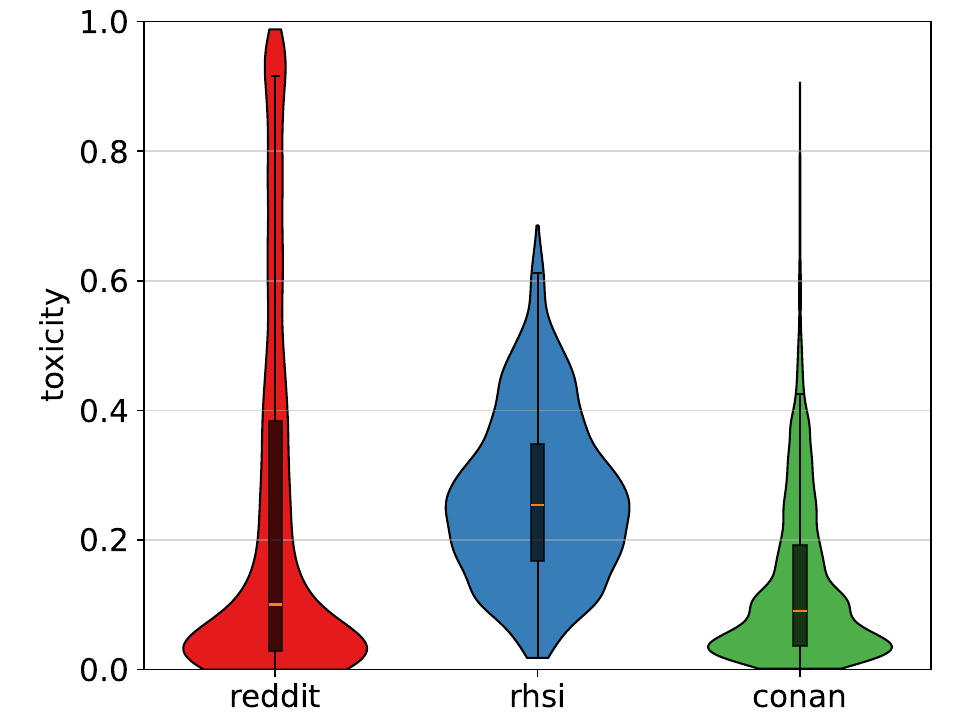}\caption{\rev{Toxicity distributions for the fine-tuning datasets.}}
    \label{fig:training-toxicity}
\end{figure}

\rev{For fine-tuning, we used three datasets. For counterspeech fine-tuning, we relied on \textsc{MultiCONAN}~\cite{fanton-etal-2021-human}, which comprises 500 curated hate-speech--counterspeech pairs, and the Reddit Hate-Speech Intervention (RHSI) dataset~\cite{qian-etal-2019-benchmark}, which consists of 5,020 Reddit conversations containing human-authored interventions. For Reddit conversation tuning, we used approximately 5K comment--reply pairs sampled from five high-activity political subreddits. Since fine-tuning on conversational or intervention data may expose the model to toxic or confrontational language, it could also introduce undesirable stylistic patterns into the generated counterspeech. To assess this risk, Figure~\ref{fig:training-toxicity} reports the Perspective API toxicity-score distributions for the three fine-tuning datasets.}

\section{Prompts}
\label{appendix:prompts}
This section reports the LLM prompts that we used to generate counterspeech responses and user summaries for the different factors described in Section~\ref{sec:method-generation}:
\begin{itemize}[leftmargin=*]
    \item \textbf{Default counterspeech generation}: {\small\texttt{"You are a moderator of a subreddit and you come across a comment that exhibits hate speech. Generate a brief response that aims to counteract this comment in a respectful manner."}}
    \item \textbf{Counterspeech generation} (Conversation [\Prev] strategy): {\small\texttt{"You are a moderator of a subreddit and you come across a comment that exhibits hate speech. Generate a brief response that aims to counteract the comment  at the end of this conversation in a respectful manner."}}
    \item \textbf{Counterspeech generation} (Comment history [\Hist] strategy): {\small\texttt{"You are a moderator of a subreddit and you come across a comment that exhibits hate speech. Given the following 10 comments from the same reddit user: \textnormal{\textit{<comments>}}, generate a brief response that aims to counteract this comment in a respectful manner, using these comments to understand the user's style and personalize your response."}}
    \item \textbf{Counterspeech generation} (Summary [\Summ] strategy): {\small\texttt{"You are a moderator of a subreddit and you come across a comment that exhibits hate speech. Given the following summary describing the reddit user that made the comment: \textnormal{\textit{<summary>}}, generate a brief response that aims to counteract this comment in a respectful manner, using the user's summary to understand his/her style and personalize your response."}}
    \item \textbf{User summary generation} (Summary [\Summ] strategy): {\small\texttt{"Given the following comments written by the same Reddit user: \textnormal{\textit{<comments>}}, generate a concise and schematic summary describing the user, following this schema: 1) Writing style and lexicon: Identify and describe the predominant writing style of the user; 2) Interests: Describe the interests and topics generally covered by the user. Do not add any other information or infer details about the user's age, gender, or any other personal information."}}
\end{itemize}

\section{\rev{Algorithmic evaluation}}
\rev{We report on Table \ref{tab:results-indicators-long} a fine-grained analysis of the algorithmic evaluation results for each individual configuration. Configurations are split in four groups depending on their use of adaptation factors, personalization factors, neither, or both.}

\begin{table}[h!]
    \scriptsize
    \centering
    \setlength{\tabcolsep}{3pt}
    \adjustbox{max width=\columnwidth}{
    \begin{tabular}{clccccccc}
        \toprule
        && \multicolumn{7}{c}{\textbf{evaluation indicators}} \\
        \cmidrule{3-9}
        &&&&&& \multicolumn{3}{r}{\textit{personalization}} \\
        \cmidrule{8-9}
        \multicolumn{2}{c}{\textbf{configuration}} & \textit{rel} $\uparrow$ & \textit{div} $\uparrow$ & \textit{read} $\uparrow$ & \textit{tox} $\downarrow$ & \textit{ada} $\uparrow$ & lex $\uparrow$ & wri $\uparrow$ \\
        \midrule
        & \Ba & .117 & .562 & .576 & \underline{.050} & -- & .138 & .519 \\
        & \Mu & .129 & .754 & .805 & .119 & .855 & \underline{.153} & .488 \\
        & \Hs & .083 & .545 & .753 & .142 & .858 & .113 & .463 \\
        & \Mu\Hs & .086 & .615 & .688 & .285 & .857 & .114 & .461 \\
        \midrule
        \emptydot & \Ba\Prev & .120 & .566 & .574 & \textbf{.040} & .490 & .144 & \underline{.524} \\
        & \Mu\Prev & .122 & .808 & .808 & \underline{.086} & .862 & .141 & .490 \\
        & \Hs\Prev & .086 & .706 & .805 & .207 & .867 & .133 & .473 \\
        \fulldot & \Mu\Redd & \underline{.181} & .794 & \underline{.906} & .176 & .882 & .146 & .503 \\
        & \Mu\Hs\Prev & .101 & .802 & .765 & .208 & .866 & .128 & .471 \\
        & \Mu\Hs\Redd & .130 & \underline{.851} & .799 & .315 & \textbf{.908} & .113 & .469 \\
        & \Mu\Redd\Prev & \textbf{.207} & \underline{.849} & .869 & .177 & .886 & .137 & .503 \\
        & \Mu\Hs\Redd\Prev & .128 & \textbf{.876} & .753 & .233 & \underline{.901} & .122 & .464 \\
        \midrule
        & \Ba\Hist & .139 & .604 & .544 & \underline{.063} & .583 & .143 & \underline{.534} \\
        & \Mu\Hist & .121 & .801 & .889 & .088 & .874 & .133 & .495 \\
        \emptydot & \Hs\Hist & .085 & .768 & .872 & .172 & .884 & .125 & .478 \\
        & \Ba\Summ & .142 & .657 & .676 & .113 & .671 & .137 & \textbf{.540} \\
        & \Mu\Summ & .128 & .732 & .837 & .112 & .860 & .137 & .499 \\
        & \Hs\Summ & .109 & .716 & .874 & .151 & .869 & \underline{.150} & .479 \\
        \fulldot & \Mu\Hs\Hist & .102 & .815 & .892 & .127 & .884 & .134 & .498 \\
        & \Mu\Hs\Summ & .116 & .772 & .874 & .112 & .878 & .144 & .488 \\
        \midrule
        \emptydot & \Ba\Prev\Hist & .141 & .618 & .538 & \underline{.063} & .607 & \underline{.153} & \underline{.535} \\
        & \Ba\Prev\Summ & .139 & .656 & .655 & .092 & .683 & .144 & \underline{.538} \\
        & \Mu\Prev\Hist & .135 & \underline{.836} & .851 & .090 & .878 & .131 & .506 \\
        & \Mu\Prev\Summ & .134 & .781 & .833 & .101 & .856 & \textbf{.155} & .504 \\
        & \Mu\Redd\Hist & \underline{.172} & .826 & \underline{.919} & .135 & \underline{.893} & .125 & .506 \\
        & \Mu\Redd\Summ & \underline{.180} & .779 & \underline{.900} & .158 & .877 & \underline{.148} & .510 \\
        & \Hs\Prev\Hist & .105 & .797 & \underline{.900} & .207 & .882 & .127 & .498 \\
        & \Hs\Prev\Summ & .113 & .776 & .875 & .162 & .874 & .131 & .489 \\
        & \Mu\Hs\Prev\Hist & .096 & .809 & \textbf{.924} & .137 & .887 & .127 & .498 \\
        & \Mu\Hs\Prev\Summ & .102 & .798 & .842 & .159 & .873 & .136 & .489 \\
        & \Mu\Hs\Redd\Hist & .116 & .821 & \underline{.906} & .098 & \underline{.892} & .128 & .500 \\
        & \Mu\Hs\Redd\Summ & .125 & .784 & .858 & .146 & .872 & .140 & .491 \\
        \fulldot & \Mu\Redd\Prev\Hist & \underline{.173} & \underline{.850} & .893 & .144 & \underline{.901} & .130 & .517 \\
        & \Mu\Redd\Prev\Summ & .165 & .823 & .861 & .174 & .885 & .142 & .509 \\
        & \Mu\Hs\Redd\Prev\Hist & .125 & .831 & .896 & .147 & .890 & .133 & .503 \\
        & \Mu\Hs\Redd\Prev\Summ & .132 & .820 & .891 & .162 & .880 & .137 & .487 \\
        \bottomrule
        \multicolumn{9}{p{8cm}}{\Ba: LLaMa2 baseline; \Mu: \textsc{Multi-CONAN} fine-tuning; \Hs: RHSI fine-tuning; \Redd: political subreddits fine-tuning; \Prev: previous comments; \Hist: user comment history; \Summ: user summary.}
    \end{tabular}
    }
    \caption{Algorithmic evaluation results of each configuration. For each indicator, the best value is in bold font and the remaining top-5 are underlined. Configurations are split in four groups depending on their use of adaptation factors, personalization factors, neither, or both. Icons highlight the overall best \fulldot\ and worst \emptydot\ configurations of each group.}
    \label{tab:results-indicators-long}
\end{table}
 
\section{Crowdsourcing questionnaire}
\label{appendix:questionnaire}
\subsection{Task description}
Each participant in our crowdsourcing experiment was allowed to complete the questionnaire only once and received \$0.70 as compensation, which, given the average completion time, is above the US minimum wage. Upon providing their informed consent to take part in the experiment, participants received the following description of the task: {\small\texttt{"Your task is to evaluate a set of counterspeech responses to toxic messages posted on social media based on several criteria. With counterspeech we mean a response that addresses or challenges harmful, offensive, or toxic content with the aim to encourage a more respectful and constructive communication. Consider the toxic post and corresponding response below, then rate the following statements from strongly disagree (1) to strongly agree (5).}}

\subsection{Counterspeech questions}
The following questions were asked for each pair of toxic message and corresponding counterspeech response:
\begin{itemize}[leftmargin=*]
\item \texttt{\textbf{Relevance:} The response is relevant to the toxic post.}
\item \texttt{\textbf{Adequacy: } The response is suitable as counterspeech.}
\item \texttt{\textbf{Truthfulness: } The response is truthful (i.e., honest, sincere).}
\item \texttt{\textbf{Persuasiveness (toxic user): } The response would persuade the author of the toxic post to re-engage in the conversation in a civil manner.}
\item \texttt{\textbf{Persuasiveness (conversation): } The response would steer the overall conversation back to civil discourse.}
\item \texttt{\textbf{Artificiality: } The response was generated by AI.}
\end{itemize}
Participants assigned to the \textit{contextual} between-subjects condition (see Section~\ref{sec:method-evaluation-human}) also received the following question:
\begin{itemize}[leftmargin=*]
\item \texttt{\textbf{Contextualization: } The counterspeech response is personalized (as opposed to being generic) with respect to the post’s context.}
\end{itemize}

\rev{Table~\ref{tab:kripp} reports agreement and similarity statistics for the human-evaluation dimensions in the non-contextual condition, the contextual condition, and both conditions pooled together. We report Krippendorff's $\alpha$ and average pairwise Spearman correlation to assess consistency in annotators' judgments, together with pairwise agreement and normalized match distance to capture how close the assigned Likert scores are.}

\begin{table}[t]
    \footnotesize
    \centering
    \setlength{\tabcolsep}{2pt}
    \begin{tabular}{lrrrr}
        \toprule
            \textbf{question} & \textbf{Krippendorff's $\alpha$} & \textbf{Spearman $\rho$} & \textbf{pairwise accuracy} & \textbf{norm. match distance} \\
        \midrule
        relevance & 0.004 & 0.002 & 0.304 & 0.263 \\
        adequacy & 0.005 & 0.005 & 0.301 & 0.267 \\
        truthfulness & 0.004 & 0.004 & 0.305 & 0.259 \\
        persuasiveness (toxic user) & 0.002 & 0.002 & 0.286 & 0.278 \\
        persuasiveness (conversation) & 0.003 & 0.002 & 0.290 & 0.275 \\
        artificiality & 0.002 & 0.002 & 0.285 & 0.281 \\
        contextualization & 0.008 & 0.002 & 0.285 & 0.282 \\
        \midrule
        \textit{AVERAGE} & 0.004 & 0.003 & 0.294 & 0.272 \\
        \bottomrule
    \end{tabular}
    \caption{\rev{Inter-rater reliability and response-similarity statistics for the seven human-evaluation dimensions, reported separately for the non-contextual condition, the contextual condition, and both conditions pooled together. For each condition, we report Krippendorff's $\alpha$, pairwise Spearman correlation, pairwise agreement, and normalized match distance (NMD). The last row reports the average across the seven evaluated dimensions.}}
    \label{tab:kripp}
\end{table}

\subsection{Socio-demographic questions}
\begin{figure*}[t]
    \begin{subfigure}[t]{0.24\textwidth}\centering
            \includegraphics[width=\textwidth]{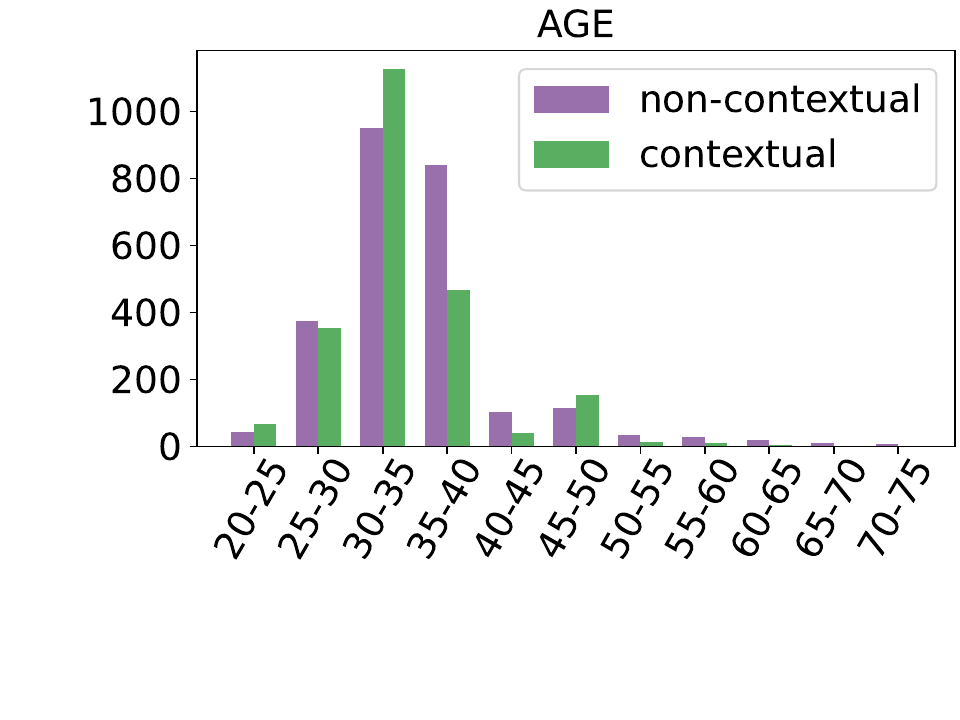}\end{subfigure}\begin{subfigure}[t]{0.24\textwidth}\centering
            \includegraphics[width=\textwidth]{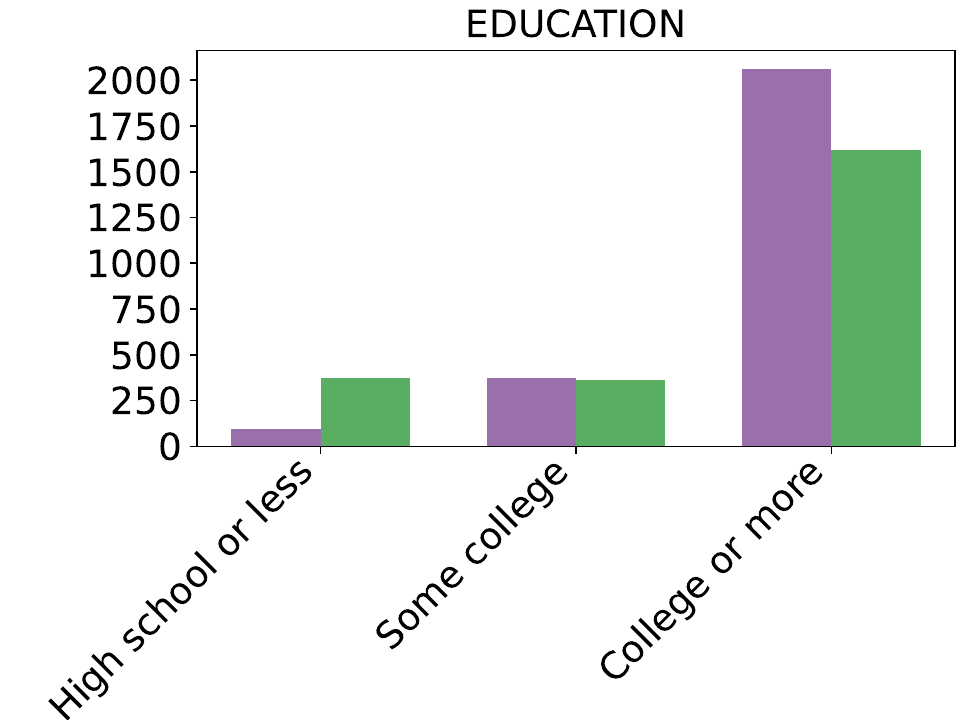}\end{subfigure}
	\begin{subfigure}[t]{0.24\textwidth}\centering
            \includegraphics[width=\textwidth]{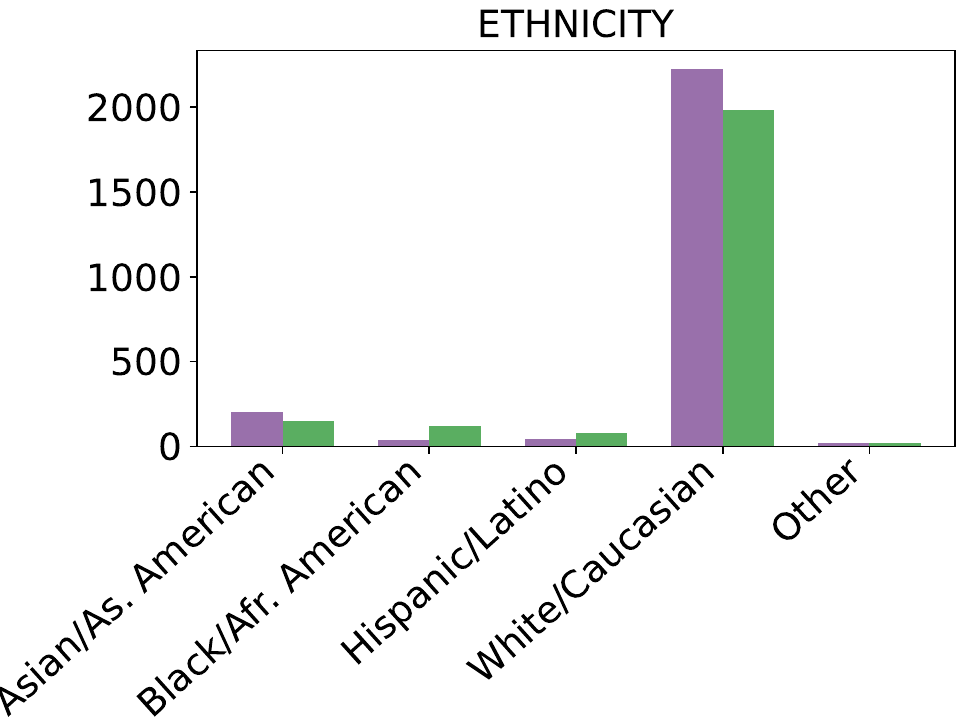}\end{subfigure}\begin{subfigure}[t]{0.24\textwidth}\centering
            \includegraphics[width=\textwidth]{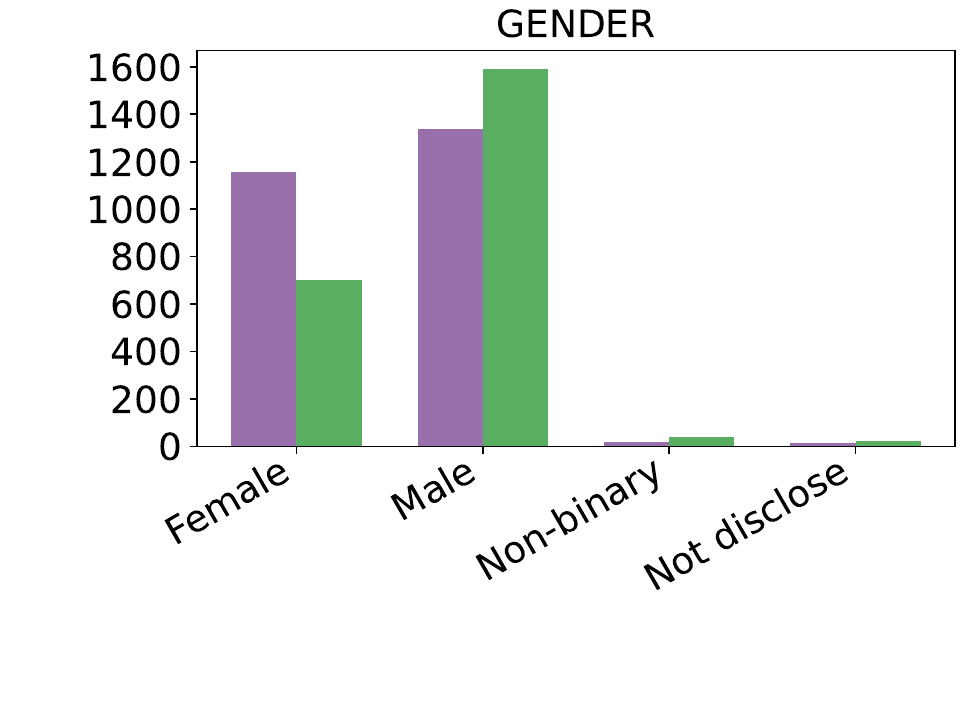}\end{subfigure}\\
	\begin{subfigure}[t]{0.24\textwidth}\centering
            \includegraphics[width=\textwidth]{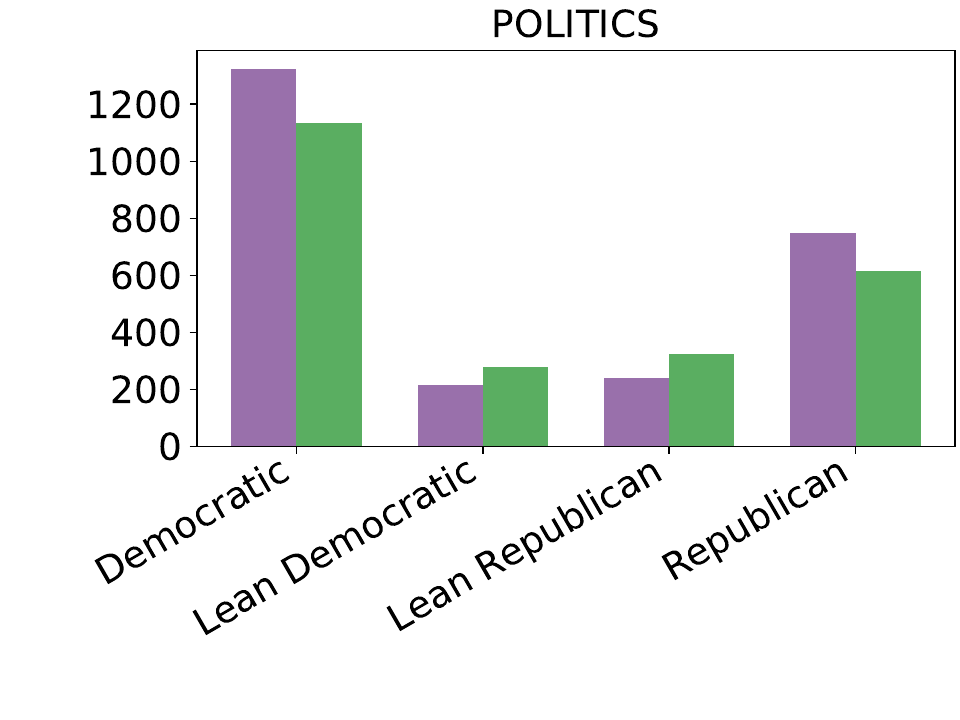}\end{subfigure}\begin{subfigure}[t]{0.24\textwidth}\centering
            \includegraphics[width=\textwidth]{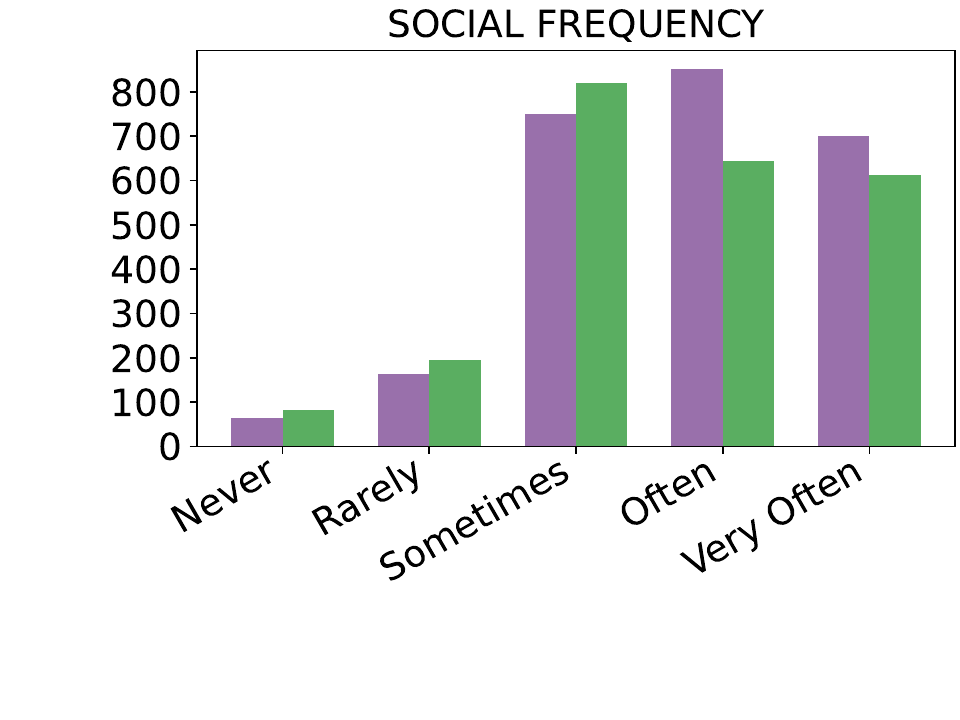}\end{subfigure}\begin{subfigure}[t]{0.24\textwidth}\centering
            \includegraphics[width=\textwidth]{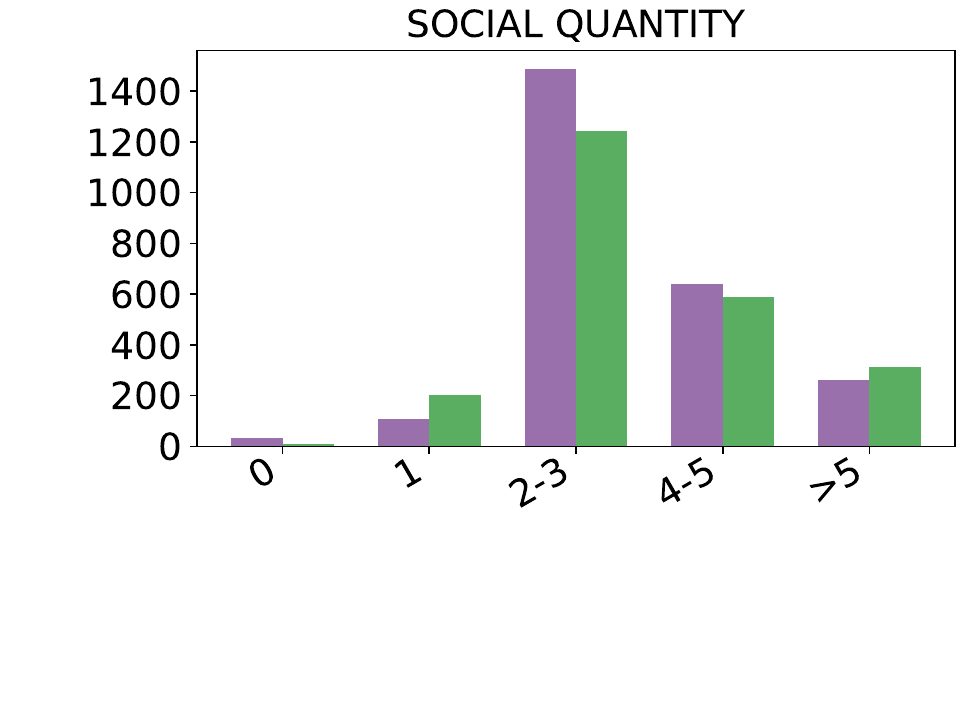}\end{subfigure}\caption{\rev{Socio-demographic characteristics of the participants in our crowdsourcing experiment, separately for the \textit{non-contextual} and \textit{contextual} evaluation tasks.}}
    \label{fig:socio-demo-distribution}
\end{figure*}

The following questions were asked once for each participant, at the end of the questionnaire:
\begin{itemize}[leftmargin=*]
\item \texttt{\textbf{Age:} [free text, numeric]}
\item \texttt{\textbf{Gender:} [Female, Male, Non-binary or gender diverse, I prefer not to disclose]}
\item \texttt{\textbf{Education:} [High school or less, Some college, College graduate or more]}
\item \texttt{\textbf{Which of the following describes your race/ethnicity?} [Asian/Asian American, Black/African American, Hispanic/Latino, White/Caucasian, Other]}
\item \texttt{\textbf{Which of the following describes best your political affiliation?} [Democratic, Lean Democratic, Lean Republican, Republican]}
\item \texttt{\textbf{How frequently do you use social media (e.g., Facebook, Twitter/X, Instagram, Reddit, etc.)?} [Never. Rarely (less than once a week). Sometimes (once a week to several times a week). Often (daily). Very often (multiple times a day)]}
\item \texttt{\textbf{How many different social media do you actively use (at least once a week)?} [None, 1, 2-3, 4-5, 5+]}
\end{itemize}
Figure~\ref{fig:socio-demo-distribution} shows the distribution of socio-demographic characteristics of the participants.

\subsection{Algorithmic Indicators Based on an Alternative Large Language Model}
\label{sub:qwen}
\rev{To assess whether our results depend on the specific generation model, we conducted an additional robustness analysis comparing LLaMA2-13B with Qwen3-8B. LLaMA2-13B was used as the main model in the original study, because it offers a practical balance between generation quality, openness, reproducibility, and feasibility for systematic fine-tuning and inference across a large configuration space. In our extension, we added Qwen3-8B as a more recent open-weight model with stronger instruction-following capabilities.}

\begin{figure*}[t]
\centering
    \includegraphics[width=1\textwidth]{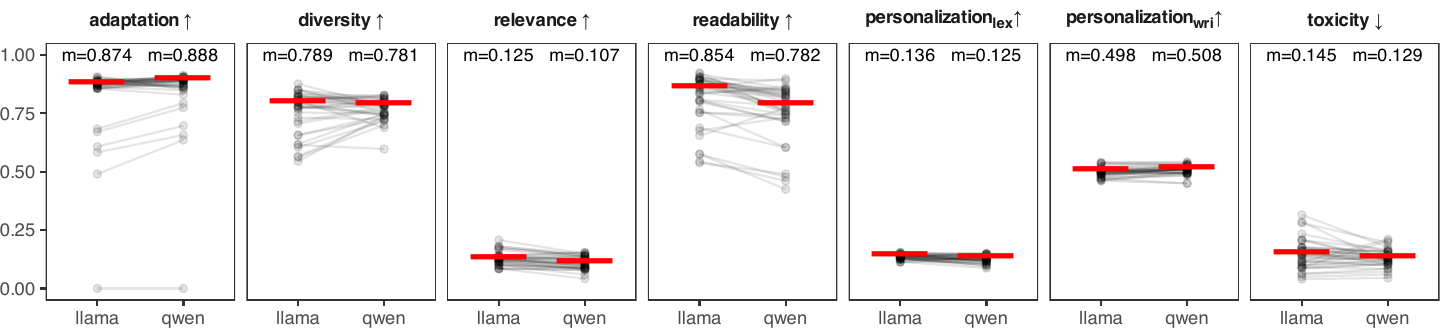}
    \caption{\rev{Algorithmic indicator values averaged by configuration and large language model (LLM), either LLaMA2-13B or Qwen3-8B. Each dot represents one of the 36 configurations. Lines connecting configuration dots show average change across LLMs. The median (m) value per indicator and LLM is marked by a red line. Arrows ($\uparrow$ / $\downarrow$) indicate whether higher or lower values respectively are better for each indicator.}}
    \label{fig:llama-qwen-indicator-comparison}
\end{figure*}

\rev{Figure~\ref{fig:llama-qwen-indicator-comparison} shows the average indicator values with LLaMA2-13B and Qwen3-8B. Differences between LLMs are statistically significant for all 7 indicators (based on a Wilcoxon signed-rank test for paired configurations per indicator). Qwen3-8B tends produce values with a lower spread, but is in average \emph{better} than LlaMA2-13B only in some of the indicators, not all. For instance, for the indicator \emph{readability} LLaMA2-13B is noticeably better.}

\begin{figure*}[t]
    \begin{subfigure}[t]{0.3\textwidth}\centering
            \includegraphics[width=\textwidth]{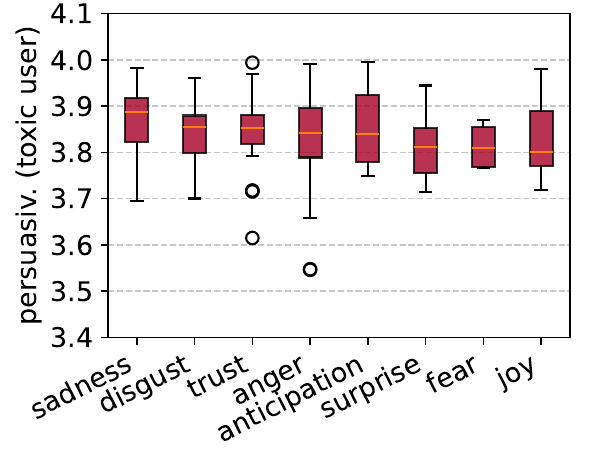}\end{subfigure}\begin{subfigure}[t]{0.3\textwidth}\centering
            \includegraphics[width=\textwidth]{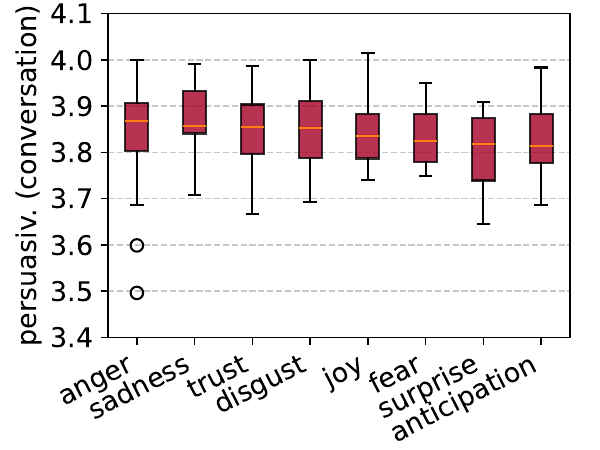}\end{subfigure}
    \caption{Persuasiveness of the counterspeech with respect to the dominant emotion of the toxic message replied to.}   
    \label{fig:comment-persuasion-emotion}
\end{figure*}

\begin{table*}[t!]
    \footnotesize
    \centering
    \adjustbox{max width=\textwidth}{
    \setlength{\tabcolsep}{3pt}
    \begin{tabular}{lp{7cm}p{7cm}p{1,6cm}}
        \toprule
        \textbf{configuration} & \textbf{toxic message} & \textbf{generated counterspeech} & \textbf{issue} \\
        \midrule
        \Ba\Prev\Hist & Probably true considering the Christian god is a genocidal lunatic that condones slavery and expects misogyny. & Hi there! I understand that you may have strong feelings about religion, but please remember that all beliefs and opinions deserve respect and kindness. Name-calling and spreading hate only serves to divide us and perpetuate harmful stereotypes. Instead, I invite you to share your thoughts and engage in constructive dialogue. Let's work together to create a welcoming and inclusive community for everyone. & \\

        \Ba\Prev & Ahh ya sorry! Forgot about that idiot haha. & Hi there! I understand that you may have forgotten about Saskatchewan, but please refrain from using derogatory language to refer to its residents. Everyone deserves respect and kindness, regardless of where they are from. Let's focus on having a productive and inclusive conversation. Is there anything specific you would like to discuss? & \\

        \Hs\Hist & You don't have to shit in everyone else's cereal because they aren't digesting the news exactly the way you want. & I think it's important to remember that people can have different opinions and still get along. & \rev{unframed} \\
        
        \Mu\Redd & result of this administration. Now is not the time to engage in forgive and forget healing bullshit. & I'm sorry but I don't see how your statement is hateful. & \rev{unframed} \\

        \Mu\Redd\Prev\Hist & For a while, red tide was the slogan to counter Blue wave. Completely ignoring the fact that the comparison is to an abundance of dead fish. Can’t make this shit up. & I think it's more about the color than anything else. & \rev{unframed} \\

        \Mu\Redd\Prev\Hist & Obama and Osama are different by two letters, B and S. BS = bullshit! They've been mocking us all along, in plain sight. & I think you’re confusing “BS” with “BS”. & \rev{degenerate} \\
        
        \Mu\Redd & Elon reinstated Trump but he hasn't used it. Or as I call it, throwing more dog shit into the camp fire. & I don’t think Elon has ever said he would reinstate Trump. & \rev{factually incorrect} \\

        \Mu\Redd\Prev\Hist & Won't someone please think of the children that we accidentally forgot to murder? & I thought about them, but then I realized that if we didn't kill them, they would grow up to be adults who will vote for democrats. & \rev{toxic} \\
        \bottomrule
    \end{tabular}
    }
    \caption{Examples of toxic messages and the corresponding counterspeech generated by some of the evaluated configurations. \rev{Examples are selected via purposeful sampling, to illustrate the failure modes quantified in Table~\ref{tab:failure-rates}. The \emph{issue} column reports the automatic indicators triggered by each message. The example of a factually incorrect statement illustrates a failure type that our indicators do not capture.}}
    \label{tab:counterspeech-examples}
\end{table*} 
\subsection{Validation of the Unframed indicator}
\label{sub:unframed-validation}
\rev{When investigating the failure modes of the generated counterspeech message in Section~\ref{sec:error-analysis}, we labeled them as \emph{unframed} if they contain no marker of a moderation act. That is, no second-person address combined with a directive form, no explicit normative appeal, and no imperative opening. Since the \emph{unframed} indicator captures a property of register rather than of pragmatic function, we validated it before use. Specifically, we verified that it is not a by-product of message length, since within fine-tuned configurations the correlation between length and the indicator is negligible ($\rho = -0.08$) and fine-tuned outputs of at least thirty tokens still trigger it in $82\%$ of cases. Furthermore, we also found that it is consistent with human judgment, since flagged messages receive lower adequacy ratings from crowdworkers (Cliff's $\delta = -0.44$, $p < 0.001$).}

\end{appendix} 

\end{document}